\documentclass[11pt,leqno]{article}
\usepackage{amsmath,amsthm,amsfonts,amssymb}

\newcommand{\vp}{\varphi}

\newcommand{\CalT}{\mathcal{T}}

\newcommand{\diag}{\text{\rm diag}}

\newcommand{\eps}{{\varepsilon}}

\newcommand{\C}{{\mathbb C}}

\newcommand{\RR}{{\mathbb R}}

\newcommand{\CC}{{\mathbb C}}

\newcommand\cB{{\cal  B}}

\newcommand\cH{{\cal  H}}

\newcommand\cI{{\cal  I}}

\newcommand\cG{{\cal  G}}

\newcommand\cP{{\cal  P}}

\newcommand\cO{{\cal O}}

\newcommand\cT{{\mathcal T}}
\newcommand\cS{{\mathcal S}}

\newcommand{\bR}{\mathbb{R}}

\newcommand{\bC}{\mathbb{C}}

\newcommand{\uz}{{\underline z}}

\def\eps{\epsilon }

\newcommand\adots{\mathinner{\mkern2mu\raise1pt\hbox{.}
\mkern3mu\raise4pt\hbox{.}\mkern1mu\raise7pt\hbox{.}}}

\newcommand\br{\begin{remark}}
\newcommand\er{\end{remark}}
\newcommand\bp{\begin{pmatrix}}
\newcommand\ep{\end{pmatrix}}
\newcommand\be{\begin{equation}}
\newcommand\ee{\end{equation}}
\newcommand\ba{\begin{equation}\begin{aligned}}
\newcommand\ea{\end{aligned}\end{equation}}

\newcommand{\bap}{\begin{app}}
\newcommand{\eap}{\end{app}}
\newcommand{\begs}{\begin{exams}}
\newcommand{\eegs}{\end{exams}}
\newcommand{\beg}{\begin{example}}
\newcommand{\eeg}{\end{exaplem}}
\newcommand{\bpr}{\begin{proposition}}
\newcommand{\epr}{\end{proposition}}
\newcommand{\bt}{\begin{theorem}}
\newcommand{\et}{\end{theorem}}
\newcommand{\bc}{\begin{corollary}}
\newcommand{\ec}{\end{corollary}}
\newcommand{\bl}{\begin{lem}}
\newcommand{\el}{\end{lem}}
\newcommand{\bd}{\begin{definition}}
\newcommand{\ed}{\end{definition}}
\newcommand{\brs}{\begin{remarks}}
\newcommand{\ers}{\end{remarks}}

\newtheorem{theo}{Theorem}[section]
\newtheorem{prop}[theo]{Proposition}
\newtheorem{cor}[theo]{Corollary}
\newtheorem{lem}[theo]{Lemma}
\newtheorem{defn}[theo]{Definition}
\newtheorem{ass}[theo]{Assumption}

\newtheorem{rem}[theo]{Remark}

\newtheorem{exams}[theo]{Examples}

\newtheorem{nota}[theo]{Notations}

\newtheorem{definition}[theo]{Definition}

\numberwithin{equation}{section}
 \title{The Erpenbeck high frequency instability theorem for ZND detonations}
\author{\sc \small
Olivier Lafitte\thanks{Universit\'e de Paris 13, LAGA and CEA Saclay, DM2S},
Mark Williams\thanks{University of North Carolina;
williams@email.unc.edu.;
Research of M.W. was partially supported by
NSF grants number DMS-0701201 and DMS-1001616.},
Kevin Zumbrun\thanks{Indiana University, Bloomington, IN 47405;
kzumbrun@indiana.edu:
Research of K.Z. was partially supported
under NSF grants no. DMS-0300487 and DMS-0801745.}}

\begin{document}

\maketitle

\begin{abstract}

The rigorous study of spectral stability for strong detonations  was begun by J.J. Erpenbeck in [Er1].  Working with  the Zeldovitch-von Neumann-D\"oring (ZND) model, which assumes a finite reaction rate but ignores effects like viscosity corresponding to second order derivatives, he used a normal mode analysis to define a stability function $V(\tau,\eps)$ whose zeros in $\Re
\tau>0$ correspond to multidimensional perturbations of a steady detonation profile that grow exponentially in time.  Later in a remarkable paper [Er3] he provided strong evidence, by a combination of formal and rigorous arguments, that for certain classes of steady ZND profiles, unstable zeros of $V$ exist for  perturbations of sufficiently large transverse wavenumber $\eps$, even when the von Neumann shock, regarded as a gas dynamical shock, is uniformly stable in the sense  defined (nearly twenty years later) by Majda.  In spite of a great deal of later numerical work devoted to computing the zeros of $V(\tau,\eps)$, the paper \cite{Er3} remains the only work we know of that presents a detailed and convincing theoretical argument for detecting them.

The analysis in [Er3] points the way toward, but does not constitute, a mathematical proof that such unstable zeros exist.    In this paper we identify the mathematical issues left unresolved in [Er3] and provide proofs, together with certain simplifications and extensions, of the main conclusions about stability and instability of detonations contained in that paper.
 %in two of the three cases he considered; namely when the quantity $c_0^2-u^2$, evaluated on the steady profile, either decreases %or increases monotonically with $x$, the distance from the von Neumann shock.   These arguments also prove a part of his result %in the case where $c_0^2-u^2$ has a single interior maximum.  Here $c_0(x)$ is the sound speed frozen at $x$ and $u(x)$ is mass %velocity relative to the von Neumann shock.
 The main mathematical problem, and our principal focus here, is to determine the precise asymptotic behavior as $\eps\to \infty$ of solutions to a linear system of ODEs in $x$, depending on $\eps$ and a complex frequency $\tau$ as parameters, with turning points $x_*$ on the half-line $[0,\infty)$.

%We expand and elucidate the argument of Erpenbeck for his
%landmark result characterizing multi-dimensional stability/instability
%of ZND detonations in the high-frequency limit.
%At the same time, we complete this argument by a rigorous
%treatment of the interval from $x_1$ to $+\infty$, missing
%in the original treatment of Erpenbeck, using a variable-coefficient
%gap lemma/repeated diagonalization argument similar to but
%somewhat more complicated than the one used in \cite{Z2} to
%treat the one-dimensional case.

\end{abstract}

%%%%%%%%%%%%%%%%%

\tableofcontents

\section{Introduction}

\emph{\quad} The most commonly studied model of combustion is the Zeldovitch-von Neumann-D\"oring (ZND) system \eqref{ZND}, which couples the compressible Euler equations for a reacting gas (in which pressure and internal energy are allowed to depend on the mass fraction $\lambda$ of reactant) to a reaction equation that governs the finite rate at which $\lambda$ changes.\footnote{Without changing the analysis below one can allow $n$ reactants, in which case $\lambda$ is an $n$-vector; for convenience we take $n=1$.}  In three space dimensions with coordinates $(x,y,z)$ a steady planar \emph{strong detonation profile} is a weak solution of this system depending only on $x$ with a jump (the stationary von Neumann shock) at $x=0$.   Without loss of generality we study profiles of the form $w=(v,u,0,0,S,\lambda)$, where $v$ is specific volume, $u>0$ is the $x$-component of particle velocity, and $S$ is entropy.  The solution is constant and supersonic ($u>c_0$, where $c_0$ is the sound speed at $x$) in $x<0$, the quiescent zone, and satisfies a nonlinear system of ODEs in the subsonic reaction zone $x>0$.  In order to be a weak solution in a neighborhood of $x=0$ it must satisfy an appropriate Rankine-Hugoniot condition at $x=0$.  There is a well-defined limiting state $w_\infty=\lim_{x\to \infty}w(x)$, which represents a state of chemical equilibrium, and the range of $u$ on $[0,\infty)$ is a compact subinterval of $x>0$.

    If one perturbs this solution, say at time $t=0$, with a multidimensional perturbation $\dot w(x,y,z)$, the profile $w(x)$ and the front defining the von Neumann shock will change with time, their description at time $t$ now being given by functions $W(t,x,y,z)$ and $x=\psi(t,y,z)$ satisfying the ZND system, a Rankine-Hugoniot condition, and the initial conditions
\begin{align}\label{a1}
W|_{t=0}=w(x)+\dot w(x,y,z),\;\psi|_{t=0}=0.
\end{align}
To a linear approximation the evolution of the perturbation is governed by the linearization of the ZND system with respect to both $W$ and $\psi$ about the steady profile $w(x)$ and steady front $x=0$.\footnote{In \cite{Er3} and in his work on non-reactive shocks \cite{Er4}, Erpenbeck  appears to have been the first to recognize the importance of linearizing  with respect to both the state $W$ and the front $\psi$.}  General perturbations can be represented by Laplace-Fourier transform in $(t,y,z)$ as superpositions of oscillatory ones, so it is natural to look for normal mode solutions of the linearized problem of the form
\begin{align}\label{a2}
\dot W(t,x,y,z)=e^{t\tau+i\alpha y+i\beta z}\phi(x,\tau,\alpha,\beta), \;\dot \psi(t,y,z)=e^{t\tau+i\alpha y+i\beta z}\xi(\tau,\alpha,\beta),
\end{align}
where $(\alpha,\beta)\in\bR^2$ and $\tau\in\bC$ with $\Re\tau\geq 0$.   Solutions \eqref{a2} with $\Re\tau>0$ and $\phi$ decaying as $x\to\pm\infty$ correspond to perturbations that grow exponentially in time.  The \emph{spectral stability problem} is to  determine whether such frequencies $(\tau,\alpha,\beta)$ exist, and if so, to locate them as well as possible.

The rigorous study of the spectral stability problem for ZND detonations was begun by Erpenbeck in \cite{Er1}. Instead of working with the $6\times 6$ system of equations satisfied by the functions $\phi=(\dot v,\dot u,\dot u_y,\dot u_z,\dot S,\dot \lambda)$ and $\xi$  in \eqref{a2}, he studied the reduced $5\times 5$ system satisfied by $\xi$ and the new unknown
\begin{align}\label{a3}
\tilde\phi:=(\dot v,\dot u,\dot U,\dot S,\dot \lambda)
\end{align}
where\footnote{Thus, $\eps/2\pi$ is the transverse wavenumber.}
\begin{align}\label{a4}
\dot U(x,\tau,\alpha,\beta):=(\alpha \dot u_y +\beta \dot u_z)/\eps, \text{ and }\eps=\sqrt{\alpha^2+\beta^2}.
\end{align}
and the other components of $\tilde \phi$  are exactly as before.

 The coefficients of the reduced system  depend only on $(x,\tau,\eps)$ so, dropping the tilde on $\phi$, Erpenbeck writes the unknowns as $\phi(x,\tau,\eps)$ and  $\xi(\tau,\eps)$.
The $5\times 5$ system of ODEs on $x\geq 0$ satisfied by the functions $\phi$ and $\xi$, together with the linearized jump condition, are given in formula (4.1) of \cite{Er1}.   The equation on $x\geq 0$ is
\begin{align}\label{a5}
\frac{d\phi}{dx}=P(x,\tau,\eps)\phi+f(x,\tau,\eps)
\end{align}
where the exact form of $f$ need not be specified now and
\begin{align}\label{a6}
P(x,\tau,\eps)=-A_x^{-1}(x)[\tau I+i\eps A_y(x)+B(x)],
\end{align}
for matrices $A_x$, $A_y$ and $B$ given in \eqref{A1}.  The $x$-dependence of these matrices enters entirely through the profile $w(x)$.

In  [Er1] Erpenbeck defined a stability function $V(\tau,\eps)$ \eqref{V} whose zeros in the right half plane $\Re\tau>0$ (``unstable zeros") correspond to perturbations of the steady profile $w(x)$ that grow exponentially in time.  Later in a remarkable paper [Er3] he provided strong evidence, by a combination of formal and rigorous arguments, that for certain classes of steady ZND profiles, unstable zeros of $V$ exist for  perturbations of sufficiently large transverse wavenumber $\eps$, even when the von Neumann shock, regarded as a gas dynamical shock, is uniformly stable.
Much numerical work has been devoted to locating zeros of $V$; see, for example, \cite{Er2,BZ2,LS,Sh,SS,HuZ} and references therein.
The argument of \cite{Er3}, though mathematically incomplete and at times incorrect in ways that we describe below, is the only convincing theoretical argument we know of that actually detects and locates unstable zeros.  Our goal here is to provide a mathematically rigorous proof of the instability (and stability) results of $\cite{Er3}$.    The main mathematical problem is to determine the precise asymptotic behavior as $\eps\to \infty$ of solutions to a linear system of ODEs in $x$ depending on $\eps$ and the complex frequency $\tau$ as parameters, with turning points $x_*$ on the half-line $[0,\infty)$.

  The computation of $V$ requires the evaluation within the reaction zone of the solution $\theta(x,\tau,\eps)$ of the homogeneous transposed equation
\begin{align}\label{a7}
\frac{d\theta}{dx}=-P^t(x,\tau,\eps)\theta\text{ on }x\geq 0
\end{align}
which satisfies the condition that $\theta$ remains bounded for fixed $(\tau,\eps)$ with $\Re\tau\geq 0$ as $x\to \infty$.  As we will see, this condition determines $\theta$ uniquely up to a constant multiple; morever, for $\Re\tau>0$ the solution $\theta$ decays exponentially to zero as $x\to\infty$.

As in \cite{Er3}  we decompose $\tau$ as
\begin{align}\label{a8}
\tau=\zeta\eps+\nu,
\end{align}
where $\zeta\in\{z\in\bC:\Re z\geq 0\}$,  $\nu\in\{z\in\bC:\Re z\geq 0, |z|\leq R\}$, and $\eps>0$ is large. In much of what follows $\nu$ will play no essential role and one can take $R=0$. However, in cases where unstable zeros of $V(\tau,\eps)$ do exist, the freedom to vary the parameter $\nu$ is useful in proving the existence of those zeros, and for that one must take $R$ sufficiently large.\footnote{Since $\eps=|(\alpha,\beta)|$, one must allow $\zeta$ to vary over all of $\Re\zeta\geq 0$ to cover all possible triples $(\tau,\alpha,\beta)$ under consideration.  The minimum size of $R$ necessary for detecting zeros of $V$ is given in part (d) of Theorem \ref{instability}. } Thus we can rewrite
equation \eqref{a7} as
\begin{align}\label{a9}
\begin{split}
&\frac{d\theta}{dx}=(\eps\Phi_0+\Phi_1)\theta\text{ where }\\
&\Phi_0(x,\zeta)=\{A_x^{-1}(x)\cdot (\zeta I+iA_y(x))\}^t\\
&\Phi_1(x,\nu)=\{A_x^{-1}(x)\cdot \nu I\}^t + \{ A_x^{-1}(x)B(x)\}^t.
\end{split}
\end{align}

The eigenvalues of the $5\times 5$ matrix $\Phi_0$ \eqref{phi0} play a crucial role in all that follows.  They are
\begin{align}\label{a10}
\mu_1=-\kappa(\kappa\zeta+s)/\eta u,\;\;\;\mu_2=-\kappa(\kappa\zeta-s)/\eta u,\;\;\; \mu_3=\mu_4=\mu_5=\zeta/u,
\end{align}
where with $c_0^2=-v^2p_v(v,S,\lambda)$
\begin{align}\label{a11}
s(x,\zeta)=\sqrt{\zeta^2+c_0^2\eta}, \;\;\;\kappa(x)=\sqrt{1-\eta}=u/c_0.
\end{align}
Here the square root defining $s$, regarded as a function of $\zeta$, is taken to be the positive branch with branch cut the segment  $[-ic_0\sqrt{\eta},ic_0\sqrt{\eta}]$ on the imaginary axis. Thus, in particular, we have
\begin{align}\label{a11z}
\begin{split}
&s=|s| \text{ when } \zeta^2+c_0^2\eta>0\\
&s=i|s|\text{ when } \zeta^2+c_0^2\eta< 0\text{ and }\zeta=i|\zeta|\\
&s=-i|s|\text{ when } \zeta^2+c_0^2\eta< 0\text{ and }\zeta=-i|\zeta|.
\end{split}
\end{align}
One checks that only $\mu_1$ has, for $\Re\zeta>0$, negative real part.    The corresponding eigenvectors are the respective columns of the matrix
\begin{align}\label{a12}
T(x,\zeta)=\begin{pmatrix}\frac{ms}{\kappa u}&-\frac{ms}{\kappa u}&-\frac{im}{1-\eta}&0&0\\\frac{\zeta}{u}&\frac{\zeta}{u}&i&0&0\\-i&-i&\frac{\zeta}{u}&0&0\\\frac{-\kappa p_S s}{um}&\frac{\kappa p_S s}{um}&0&1&0\\\frac{-\kappa p_\lambda s}{um}&\frac{\kappa p_\lambda s}{um}&0&0&1\end{pmatrix}
\end{align}
where $m=\frac{u}{v}$ is the mass flux.

On any subinterval $\cI$ of $[0,\infty)$ where (for fixed $\zeta$) the transformation $T(x,\zeta)$ is invertible, we  set $\theta=T(x,\zeta)\pi$ and obtain the system with diagonal leading term
\begin{align}\label{a12a}
\begin{split}
&\frac{d\pi}{dx}=(\eps D+E)\pi,\text{ where }\\
&D:=\diag(\mu_1,\dots,\mu_5)\text{ and }E(x,\zeta,\nu):=T^{-1}\Phi_1 T-T^{-1}\frac{dT}{dx}.
\end{split}
\end{align}
On such subintervals one constructs\footnote{See for example Chapters 5 and 6 of Coddington and Levinson \cite{CL}.}
%CHANGED: more clear.KZ
%approximate solutions
approximate solutions to order $\eps^{-m}$
%ENDCHANGED
of \eqref{a12a} associated to each of the eigenvalues $\mu_i$ of the form
\begin{align}\label{a13}
\pi_i(x,\tau,\eps)=e^{\eps h_i(x,\zeta)+k_i(x,\zeta,\nu)}
[f_{i0}(x,\zeta,\nu)+\eps^{-1}f_{i1}(x,\zeta,\nu)+\dots
%CHANGED: same typo noted earlier.-KZ
%+\eps^{-m}f_{im}(x,\zeta,\nu)],
+\eps^{-(m+1)}f_{i(m+1)}(x,\zeta,\nu)],
%ENDCHANGED
\end{align}
where
\begin{align}\label{a14}
h_i(x,\zeta)=\int^x_0\mu_i(x',\zeta)dx'
\end{align}
and $f_{i,0}$ is the unit vector $e_i$ with $i$-th component $1$. This is done by substituting the expansion into \eqref{a15} and solving the equations obtained by equating coefficients of equal powers of $\eps$.  For $i=1,2$ we will use the formulas
\begin{align}\label{a15}
k_i(x,\zeta,\nu)=\int_{0,C_w}^xE_{ii}(x',\zeta,\nu)dx'
\end{align}
where $C_w$ is a contour that lies on the real axis, except for short excursions into the upper or lower half plane to avoid singularities of $E_{ii}$ \eqref{E11}.  The choice of $C_w$ depends on the profile $w$ and is explained in sections \ref{matchI} and
\ref{matchD}.  We do not need explicit formulas  for the other quantities appearing in \eqref{a13} for $i=1,\dots,5$.   Sufficient conditions for approximate solutions of this type to be
%CHANGED
%close to true exact solutions
$O(\eps^{-(m+1)})$ close
in relative error
%ENDCHANGED
to true exact solutions
of \eqref{a12a} for $\eps$ large  are given in Theorem \ref{mpp}.

From the formulas \eqref{a10} we see that for any fixed value of $\zeta$, the eigenvalues $\mu_1$ and $\mu_2$ are distinct except at $x$ values where $s^2=\zeta^2+c_0^2\eta=0$; at such values the first and second columns of $T$ are parallel.  The eigenvalues $\mu_2$ and $\mu_3$ are distinct except at $x$ values where $\zeta=u$, and then the second and third rows of $T$ are clearly parallel.   For all other values of $x$ the matrix $T(x,\zeta)$ is invertible.  The special values $x_*=x_*(\zeta)$ where $T$ is singular are referred to as ``turning points".  These points determine a family of subintervals $\cI\subset [0,\infty)$, and on each of these subintervals the approximate solutions described above can be constructed.\footnote{The family of subintervals depends on $\zeta$, but we normally suppress  $\zeta$ in denoting both the intervals and the $x_*$.}

A complex number $\zeta$ with $\Re\zeta\geq 0$ is defined in \cite{Er3} to be of Class III or Class II respectively, when there exists $x_*\in [0,\infty]$ such that
$s(x_*,\zeta)=0$ or $\zeta=u(x_*)$.\footnote{When $x_*=\infty$, the obvious limits are intended here.  For a given $\zeta\in\text{Class III or Class II}$,  there may be more than one $x_*$ such that $s(x_*,\zeta)=0$ or $\zeta=u(x_*)$.}  All other $\zeta$  are said to be of Class I.  Thus we have
\begin{align}\label{a16}
\begin{split}
&\text{Class III}  = \{\zeta:\Re\zeta=0\text{ and }\min_x (c_0\eta^{\frac{1}{2}})\leq |\zeta|\leq \max_x (c_0\eta^{\frac{1}{2}})\}\\
&\text{Class II}= \{\zeta:\Im \zeta=0\text{ and }\min_x u\leq \zeta\leq \max_x u\}\\
&\text{Class I}=\{ \text{all remaining }\zeta\in\bC\text{ with }\Re\zeta\geq 0\}.
\end{split}
\end{align}
Class III (resp. II) consists of two (resp. one) bounded closed interval(s), and the minima appearing in \eqref{a16} are positive.

\begin{nota}
When working with Class III values of $\zeta$ we will usually suppose $\zeta=i|\zeta|$.  Similar results hold with similar proofs when $\zeta=-i|\zeta|$, but certain formulas change slightly.  Thus, for some statements it is helpful to define
\begin{align}\label{plus}
\text{Class III}_+=\text{Class III}\cap \{\zeta=i|\zeta|\}.
\end{align}

\end{nota}

A simpler form \eqref{n3} of the stability function $V(\tau,\eps)$, based on the choice of a ``good unknown" that removes the front from the linearized ZND system in the reaction zone,  was derived in \cite{CJLW}.  That form involves no integrals $b_j$ as in \eqref{V1} and was  shown in \cite{CJLW} to be equal to Erpenbeck's stability function.\footnote{ It was also shown to be a nonzero multiple of the ZND Evans function defined in \cite{JLW}.} The formula \eqref{n3} for $V(\tau,\eps)$ depends on the value at $x=0$ of the exact decaying solution $\theta(x,\tau,\eps)$ of \eqref{a7}.  Our assumptions guarantee that $\theta$ and $P^t(x,\tau,\eps)$ are real analytic in $x$ on $[0,\infty)$, so for fixed $(\tau,\eps)$ the function $\theta$ can be extended analytically as a solution of
\eqref{a7} for $z$ in some complex neighborhood of $[0,\infty)$.   For a given $\zeta$ let us suppose that there are a finite number $N$ of turning points ($N=0$ for $\zeta\in\text{Class I}$):
\begin{align}\label{a17}
0<x_{N*}<x_{N-1*}<\cdots<x_{1*},
\end{align}
The strategy of \cite{Er3} for determining $\theta(0,\tau,\eps)$ is first to determine $\theta$ on the interval $[x_{1*}+\delta,\infty)$ for some $\delta>0$, then to analytically continue that solution in $\bC\setminus x_{1*}$ around the turning point $x_{1*}$ and up the real axis to $x_{2*}+\delta$, then to analytically continue around $x_{2*}$ in $\bC\setminus x_{2*}$, and  to repeat this process until $x=0$ is reached.  It is impossible to determine explicit formulas for the exact solution $\theta$, so instead one tries to determine explicit formulas that closely approximate the exact solution  for $\eps$ large on a certain collection of subsets of complex plane, subsets whose union covers $[0,\infty)\setminus \cT$, where $\cT$ is a union of small intervals each centered at a turning point.   The approximating formula will generally change from subset to subset, so there arises a \emph{matching problem} in passing from a given subset to an adjacent one.

On \emph{bounded} subsets of the complex plane we use approximating integral formulas of type \eqref{a13} and linear combinations thereof.  Sufficient conditions for such formulas to closely approximate exact solutions of \eqref{a7} (for $\eps$ large) together with precise error estimates are given in Theorem \ref{mpp}.
We will use the notation $\bar\theta_j$ to denote an exact solution defined on some bounded region $\cO_j\subset \bC$ which lies close to an approximate solution $\theta_j=T\pi_j$ defined by a formula of type \eqref{a13} on a possibly larger region $\cO_j'$.
Theorem \ref{mpp} does not apply to unbounded regions; in order to obtain an accurate approximating formula on the unbounded interval $[x_{1*}+\delta,\infty)$ we use a completely different approach based on the variable-coefficient gap lemma, Lemma \ref{vargaplem}, introduced recently in \cite{Z2}.

Following \cite{Er3}, when considering Class III values of $\zeta$ we focus on profiles $w(x)$ for which the quantity
$c_0^2\eta(x)=c_0^2(x)-u^2(x)$ either increases monotonically for $x\in [0,\infty)$ (case I),  decreases monotonically (case D), or increases to a maximum at $x_M$ and then decreases (case M).   In case M our analysis omits treatment of the special Class III$_+$ value of $\zeta$ given by
\begin{align}\label{a22}
\zeta_M=i c_0\eta^{\frac{1}{2}}(x_M).
\end{align}
The turning point argument for such $\zeta_M$ in \cite{Er3} is much more involved than that for the other Class III values of $\zeta$; we do not discuss that case here.
The turning point arguments for other Class III values of $\zeta$ in case M, given in the proof of Proposition \ref{M}, are a combination of the arguments for cases I and D.   Erpenbeck shows on p.1303 of \cite{Er3} that for ``$A\to B$ reactions" (one reaction detonations with no back reaction) with an Arrhenius rate law \eqref{rate2}, the cases I, D, and M actually occur, the distinction depending on values of the heat capacity ratio, heat of reaction, and Mach number of the detonation. His results indicate that for $A\to B$ reactions the case of a single interior minimum does not actually occur.

The arguments of \cite{Er3} are incomplete in two ways, namely,  in the treatment of the unbounded interval $[x_{1*}+\delta,\infty)$ and in the problem of matching solutions on adjacent regions.  To treat the unbounded interval \cite{Er3} considers the limiting problem
\begin{align}\label{a18}
\frac{d\theta}{dx}=-P^t(\infty,\tau,\eps)\theta
\end{align}
obtained by evaluating the coefficients of \eqref{a7} at the values given by $\lim_{x\to \infty}w(x)$,
and refers to an argument of \cite{Er1} showing that for large $x$, solutions of \eqref{a7} lie close to solutions of \eqref{a18}.  However, \cite{Er1} is not concerned with behavior as $\eps\to \infty$, and this argument does not address the possibility that as $\eps$ increases  solutions of \eqref{a7} remain close to solutions of the limiting problem \eqref{a18} only on intervals $[x(\eps),\infty)$ such that $\lim_{\eps\to\infty}x(\eps)=\infty$.   Too rapid growth of $x(\eps)$, for example $O(e^{C\eps})$, would make it impossible to carry out the matching arguments for the remaining interval.

To obtain an approximating formula for the decaying solution on the unbounded interval, \cite{Er3}  observes that the integrals defining $h_1(x,\zeta)$ and $k_1(x,\zeta,\nu)$ ``become linear in $x$ near the region of chemical equilibrium", i.e., as $x\to \infty$.
%Since the integrals defining the corresponding exponents in the decaying solution of \eqref{a18} are linear,
Since for most choices of $\zeta$,  the eigenvalue $\mu_1(x,\zeta)$,  the corresponding eigenvector of $\Phi_0(x,\zeta)$, and $E_{11}(z,\zeta,\nu)$ approach their counterparts, which are constant in $x$, in the perturbation expansion for $\eps$ large of the eigenvalues and eigenvectors of $\eps\Phi_0(\infty,\zeta)+\Phi_1(\infty,\nu)$, \cite{Er3} concludes that ``the $\eps\to \infty$ limit and the $x\to \infty$ limit are interchangeable", and that on the unbounded interval the exact decaying solution is well-approximated by the approximate solution\footnote{Erpenbeck writes ``$\theta\sim\theta_1$" for $x>x_{1*}$.
%The symbol $\sim$, which is never clearly defined in \cite{Er3}, is meant to indicate that for $\eps$ large the functions are %close on the set of $x$ indicated.
According to the standard definition in asymptotic ODE theory, given in \cite{CL} for example,
\begin{align}\label{a19}
\theta\sim\theta_1 \text{ for }x\in S\Leftrightarrow \text{ there exist  positive constants }C, R \text{ such that }|\theta-\theta_1|\leq \frac{C}{\eps}|\theta_1|\text{ for }x\in S, \eps >R.
\end{align}
}
\begin{align}\label{a20}
\theta_1(x,\tau,\eps)=T(x,\zeta)\pi_1(x,\tau,\eps).
\end{align}
He then uses $\theta_1$ in the matching arguments on the remaining finite interval.

This argument is suggestive but nonrigorous, and it actually leads to an incorrect conclusion.  In
Theorem \ref{conjugation} we use an argument based on multiple conjugations and the variable coefficient gap lemma, Lemma \ref{vargaplem}, to construct the exact decaying solution $\theta$ on the unbounded interval.  The construction provides at the same time an accurate approximating formula and an error estimate (see (2.9)). Comparison of this formula with $\theta_1$ shows that  the solution
$\theta(x,\tau,\eps)$ differs from $\theta_1(x,\tau,\eps)$ by a complex factor (namely $e^{H_\eps(x,\zeta,\nu)}$ in \eqref{3dd}) that depends on $x$ and  grows \emph{exponentially} with $x$ for certain choices of $(\zeta,\nu)$.  Thus, it is not true that $\theta\sim\theta_1$ (see the preceding footnote);  in fact $\theta\sim M(\tau,\eps)\theta_1$  is not true on any open $x$-interval for any multiple $M(\tau,\eps)$ of $\theta_1$.    Even for choices of $(\zeta,\nu)$ for which the factor $e^{H_\eps(x,\zeta,\nu)}$ remains bounded as $x\to\infty$, the presence of this factor means that the matching argument for the unbounded interval and its adjacent bounded region has to be done with a carefully chosen multiple of $\theta_1$ (or alternatively, of $\theta$) and not $\theta_1$ itself.

%The matching arguments of \cite{Er3} are incomplete also in that they neglect to keep track of the grow

In the matching arguments one uses bases of exact solutions $\{\bar\theta_1,\dots,\bar\theta_5\}$ of \eqref{a7} whose asymptotic behavior as $\eps\to \infty$ is known in regions of the complex plane that include portions of the adjacent intervals under consideration.
In these arguments one must take into account the growth rate with respect to $\eps$ not only of the solutions $\bar\theta_j$ (to know, for example, which solutions are ``dominant" or ``recessive" in certain regions), but also the growth rate of coefficients.
For example, if $\theta$ denotes the exact decaying solution on the interval $[x_{1*}+\delta,\infty)$ constructed by Lemma \ref{vargaplem}, the first step in determining the analytic continuation of this solution around $x_{1*}$ is to expand $\theta$ in such a basis:
\begin{align}\label{a21}
\theta(z,\tau,\eps)=c_1(\tau,\eps)\bar\theta_1(z,\tau,\eps)+\dots+c_5(\tau,\eps)\bar\theta_5(z,\tau,\eps).
\end{align}
The coefficients in \eqref{a21} are independent of $z$ but depend on $(\tau,\eps)$, and a necessary part of a rigorous matching argument, a part that is missing in \cite{Er3}, is to keep close track of the growth rates of these coefficients with $\eps$ as one passes from one basis to another.  This type of analysis is carried out in the proofs of Propositions \ref{88} and \ref{44l}.

In case D  we provide some arguments that seem to be missing in the analysis of \cite{Er3}.  As in case I, where there is also just a single turning point $x_*$ for class III values of $\zeta$, after the decaying solution $\theta$ has been analytically continued just a little to the left of $x_*$, there remains the question of determining the asymptotic behavior as $\eps\to \infty$ of the solution on the remaining part of the interval $[0,x_*-\delta]$.  That determination requires knowledge of the asymptotics of all  the exact solutions in the basis used to express $\theta$ \eqref{8e},\eqref{8j}.  The approach of \cite{Er3} is to use the method of the parameter problem, Theorem \ref{mpp}, for this purpose.  That approach works only in case I, where $\Re\mu_j=0$ on $[0,x_*-\delta]$; in case D it turns out to be impossible to use Theorem \ref{mpp} to determine the asymptotic behavior of all but one of the basis solutions on $[0,x_*-\delta]$.
The reason, explained in step \text{6} of the proof of Proposition \ref{44l}, is that certain differences $\Re(\mu_i-\mu_j)$ have an unfavorable sign on $[0,x_*-\delta]$.   In step \text{6} an energy-type estimate is provided along with an analysis of coefficients to complete the treatment of this case.  Such arguments are needed as well in case M, where there may be two turning points associated to a given $\zeta\in \text{Class III}$, for understanding the asymptotic behavior of the exact solution on the full interval between the turning points.

To treat Class III values of $\zeta$ that correspond to three or more turning points, matching arguments beyond those given in this paper may be needed, particularly for the continuation around $x_{k*}$ when $x_{(k-1)*}$ is of increasing type.

\subsection{Assumptions}
\begin{ass}\label{thermo}
The thermodynamic functions appearing in the ZND system \eqref{ZND}, $p$ (pressure), $T$ (temperature), $\Delta F$ (free energy increment), and $r$ (reaction rate) are real analytic functions of their arguments $(v,S,\lambda)$.
\end{ass}

\begin{ass}\label{profile}
The steady strong detonation profile $w(x)=(v,u,0,0,S,\lambda)$ is a real-analytic function of $x$ in the subsonic reaction zone $[0,\infty)$.   There exist constants $C_1$ and $C_2$ such that
\begin{align}
0<C_1\leq \kappa=\frac{u}{c_0}\leq C_2 <1 \text{ and } 0<C_1\leq {u}\leq C_2 <1 \text{ for all }x\in [0,\infty).
\end{align}
Moreover, there exist positive   constants $C$, $\beta$ such that
\begin{align}\label{profile2}
|w(x)-w(\infty)|\leq C e^{-\beta x}.
\end{align}
We mainly study profiles of type I, D, or M, and will call attention to results that hold without this restriction.\footnote{For $\zeta$ of Class III there is a single turning point $x_*$ in cases I and D.  In case M, depending on the choice of $\zeta\in \text{Class III}$, there can be one or two turning points.}

\end{ass}

\begin{ass}\label{rate}
The rate function satisfies
\begin{align}\label{a22c}
r|_{\lambda=0}=0,\;\;r_\lambda < 0,\; r_v|_{\lambda=0}=0,\; r_S|_{\lambda=0}=0.
\end{align}
\end{ass}

This assumption is satisfied, for example, by rate functions of the form
\begin{align}\label{rate1}
r=-k\rho\phi(T)\lambda,
\end{align}
where $\rho$ is density and $k>0$ is a reaction rate constant, such as the Arrhenius rate law
\begin{align}\label{rate2}
r=-k\lambda \exp(-E/RT)\text{\;\;($E$ is activation energy)}
\end{align}
considered  in the section of \cite{Er3} on $A\to B$ detonations.

\begin{rem}\label{rate3}
\textup{Assumption \ref{rate} can be weakened to allow for rate functions (such as those incorporating back reactions $B\to A$) for which one may not have $r_v=0$ and $r_S=0$ when $r=0$ (equilibrium).  Inspection of the proof of Theorem \ref{conjugation} shows that it suffices there to have for all $x\geq 0$:
\begin{align}\label{o1}
\Re(\mu_j^*-\mu_1^*)(x,\zeta,\nu,h)\geq -\delta_*h+O(he^{-\beta x}), \;j=2,\dots,5\text{ where }\delta_*<\beta,
\end{align}
where $\beta$ is as in \eqref{profile2}, $h=1/\eps$, and the $\mu_j^*$ are the eigenvalues of $\Phi_0(x,\zeta)+h\Phi_1(x,\nu)$.
For rate functions satisfying \eqref{a22c} we check below that \eqref{o1} is satisfied with $\delta_*=0$. It is hard to check \eqref{o1} directly, but a readily verifiable condition that implies \eqref{o1} is
\begin{align}\label{o2}
\Re(\mu_j^*-\mu_1^*)(\infty,\zeta,\nu,h)\geq -\delta_*h \;j=2,\dots,5\text{ where }\delta_*<\beta,
\end{align}
the corresponding condition for the constant coefficient limiting system $\Phi_0(\infty,\zeta)+h\Phi_1(\infty,\nu)$. For such rate functions the remaining arguments of this paper are also valid (see also Remark \ref{o3}).}

%vanish at equilibrium even though $\lambda$ may not vanish at equilibrium. In such cases   It is necessary, though, to assume that... }
\end{rem}

\begin{ass}\label{vN}
The stability function for the von Neumann shock, defined originally in \cite{Er4},\footnote{This stability function turns out to agree with the function $L_1(\zeta)$ \eqref{n5}, described in section \ref{instab}, and to be a nonvanishing multiple of the Majda stability determinant for shocks defined in \cite{M}} has no zeros in $\Re\zeta\geq 0$.
\end{ass}
This means that the equation of state of the unreacted explosive is such that the von Neumann step-shock would be stable if the reactions behind it were somehow suppressed.  This assumption is made in \cite{Er3} to allow us to concentrate on instability which arises solely from the reactions.  It always holds, for example, for step-shocks in ideal polytropic gases.

\subsection{Main results}
\emph{\quad} The main results of this paper provide a rigorous determination  of $\theta(0,\tau,\eps)$ (and, in fact, of $\theta(x,\tau,\eps)$ for $x$ away from turning points)  to arbitrarily high accuracy for $\eps$ large and are contained in Theorem \ref{conjugation}, Proposition \ref{88}, and Proposition \ref{44l}.  The behavior at $x=0$ for $\eps$ large may be described informally as follows.  Let $t_1(\zeta)$ and $t_2(\zeta)$ be the first two columns of the matrix $T(0,\zeta)$ \eqref{a12}.  With $\tau=\zeta\eps+\nu$ we have
\begin{align}\label{mr}
\begin{split}
&\text{ For }\zeta \notin \text{Class III}, \;\;\theta(0,\tau,\eps)=t_1(\zeta)+O(1/\eps)\\
&\text{ For }\zeta\in \text{Class III}_{+}\setminus E, \;\; \theta(0,\tau,\eps)=t_1(\zeta)+\alpha(\eps,\zeta,\nu)t_2(\zeta)+O(1/\eps),
\end{split}
\end{align}
where $\alpha$ is given by \eqref{4u} and the exceptional set $E=\{ic_0\eta^{1/2}(0+),ic_0\eta^{1/2}(\infty)\}$ in cases I and D. In case M, $E$ also contains $ic_0\eta^{1/2}(x_M)$, where $x_M$ is the location of the maximum.

The main instability result of \cite{Er3}, restated here as Theorem \ref{instability}, follows then by an argument using Rouch\'e's Theorem and the formula \eqref{n3} for $V(\tau,\eps)$ in terms of $\theta(0,\tau,\eps)$. In cases I and M this theorem provides an explicit sufficient condition \eqref{n10z}, expressed in terms of  functions of the steady flow variables, for detecting the presence of zeros of $V(\tau,\eps)$ in $\Re\tau>0$ when $\eps$ is sufficiently large.  The unstable zeros occur for a subinterval of Class III values of $\zeta$ at certain values of $\nu$ with $\Re\nu>0$.  In case D the result allows one to conclude for all $\zeta$ \emph{except} the two particular $\text{Class III}_+$ values in \eqref{a18b}  that for $\eps\geq \eps(\zeta)$ sufficiently large, $V(\zeta\eps+\nu,\eps)$ does not have unstable zeros.  Moreover, for the $A\to B$ reactions  described above, \cite{Er3} shows that for certain values of the physical parameters, unstable zeros are actually present in cases I and M.

Our use of Lemma \ref{vargaplem} permits  certain simplifications and extensions relative to \cite{Er3}.  The application of the lemma to this problem depends only on having good separation between $\mu_1$ and the remaining set of eigenvalues.  Thus, the crossing of $\mu_2$ and $\mu_3$ that occurs for Class II values of $\zeta$ at the associated turning points are irrelevant for us.  In our analysis it is only necessary to distinguish two sets of $\zeta$, namely,  Class III and its complement in $\Re\zeta\geq 0$.
In \cite{Er3} the unbounded interval could be taken to be $[0,\infty)$ only for Class I values of $\zeta$; here we may take it to be $[0,\infty)$ for both Class I and Class II.  The special Class II values given by
\begin{align}\label{a18a}
\zeta=u(0+)\text{ and }\zeta=\lim_{x\to\infty}u(x):=u(\infty),
\end{align}
which correspond to  turning points ``at $0$ and at $\infty$",
had to be excluded from the analysis of \cite{Er3}, but they can be treated in our approach like any other Class II values.
Thus, for all \text{non-Class III} values, $\theta(0,\zeta,\nu,\eps)$ can now be determined in a single step, with no matching required.  However,
%NOT CHANGED, KZ (it is consistent with footnote and rest of
%paper, the meaning is clear, and the alternatie is awkward.)
%like \cite{Er3} we must exclude from consideration the exceptional $\text{Class III}_+$ values
like \cite{Er3} we must exclude from consideration the exceptional $\text{Class III}_+$ values
%ENDCHANGED
\begin{align}\label{a18b}
\{ic_0\eta^{\frac{1}{2}}(0+), ic_0\eta^{\frac{1}{2}}(\infty)\}
\end{align}
in the matching arguments of section \ref{match}.\footnote{Although \cite{Er3} does not explicitly exclude $ic_0\eta^{\frac{1}{2}}(0+)$, we believe that his analysis does not cover this case.}

Lemma \ref{vargaplem} also permits us to keep track of how the constants that appear in our estimates depend on $\zeta$. For example, our proof of Theorem \ref{instability} shows that for $\zeta$ in any compact subset $K$ of $\left(\{\Re\zeta\geq 0\}\setminus\text{Class III}\right)$  there exists a positive constant $\eps(K)$ such that
  $V(\tau,\eps)\neq 0$ for $\eps\geq \eps(K)$.
Obtaining uniform wavenumber cutoffs $\eps(K)$ has an obvious importance for  numerical investigations of the
presence of unstable zeros.
There was no attempt to get uniform cutoffs in \cite{Er3}, and in fact such cutoffs cannot be extracted from the arguments given there  for reasons already indicated (for example, incorrect treatment of the unbounded interval and the need to use matching arguments for Class II values of $\zeta$ due to crossing of $\mu_2$ and $\mu_3$).   Moreover, our proof of the cutoff result just stated holds even for profiles that are not of type I, D, or M; such profiles are not studied in \cite{Er3}.
We also prove statements about uniform choices of $\eps$ in the parts of Theorem \ref{instability} that pertain to Class III values and instability.

We  introduce a major simplification in the proof of Theorem \ref{instability} by working with the simpler form \eqref{n3} of $V(\tau,\eps)$, instead of with Erpenbeck's original form \eqref{n1}.  The analysis in \cite{Er3} of the asymptotic behavior of the integrals $b_1$ and $b_2$ \eqref{V1}, which is complicated since the integrands involve $\theta$, is thus no longer needed.  The use of the simpler form of $V$ also gives more rapid convergence of the approximate stability function $L_a$ to $L$ (see \eqref{n4}; here $V=\eps L$), which in turn yields wavenumber cutoffs $\eps(K)$ that are better (i.e., smaller) than those derivable using  the original form of $V$.

\subsection{Discussion and open problems}
\emph{\quad}A rigorous approach to some aspects of the high frequency behavior of the ZND stability function $V(\tau,\eps)$ was given in \cite{CJLW}.  That paper addressed the question of the behavior of the exact decaying solution $\theta$ of \eqref{a7} on the unbounded interval.  A different type of gap lemma was used there to show that $\theta(x,\tau,\eps)$ (suitably normalized) is well-approximated by the solution $\theta_L(x,\tau,\eps)$ of the limiting problem \eqref{a18} on unbounded subintervals $[x(\eps),\infty)$,  where  $x(\eps)$ converges to $\infty$, but not too fast to rule out matching arguments.  No comparison was made there between $\theta_L$ and $\theta_1$.

The present treatment based on Lemma \ref{vargaplem} provides several improvements over \cite{CJLW}.  First, we obtain an approximating formula for $\theta$ that is
valid on unbounded intervals whose left endpoint is independent of $\eps$.
In  Theorem 5.1 of \cite{CJLW} a separate argument using the Tracking Lemma of \cite{Z3} was used to prove the absence of unstable zeros for $\eps$ sufficiently large only for a certain proper subclass of non-Class III values (mostly away from $\Re\zeta=0$).  Turning points, the behavior of $V(\tau,\eps)$ for Class III values of $\zeta$, and instability were not treated in \cite{CJLW}.

\medskip
%HERE: open problems...
Our study completes the mathematical analysis of \cite{Er3},
yielding rigorous general criteria for instability of ZND detonations.
Despite extensive numerical and asymptotic studies of instability
in the detonation literature, this represents to our knowledge the
only analytic proof of instability.

Another interesting and closely related problem is to establish
conditions for {\it stability} of detonation waves in the {\it small
heat-release limit} $\Delta F\to 0$, in which equations
\eqref{ZND} formally decouple,
and the associated {\it high overdrive limit},\footnote{
As described in \cite{Z1,Z2}, this amounts to a simultaneous
large shock-strength and zero heat release/activation energy limit;
for further description, see \cite{Z2} and Appendix C, \cite{Z1}.}
as carried out in the one-dimensional case in \cite{Z2}.
As discussed in \cite{Z2}, both of these problems reduce to establishing
uniform high-frequency stability estimates similar to those obtained
in the present paper;
as noted in \cite{Er2}, the latter problem is quite subtle in multi-dimensions.
Such a result would, together with the instability results developed
in \cite{Er3} and this paper, complete the picture of behavior
that has emerged from numerical lore of transition from stability
to instability as heat release or overdrive is varied.

The study of positive high-frequency stability criteria is further motivated
by the needs of numerical stability analyses,
where, to obtain definitive results of stability,
it is necessary to perform some such asymptotic analysis
truncating the computational domain to a bounded set of frequencies.
See \cite{BZ1,BZ2,Z2} for treatments in the one-dimensional case.

What is needed beyond the analysis of the present paper to obtain
positive high-frequency stability criteria is to treat the Class III
frequencies neglected here and in \cite{Er3}
for which turning points $x_*$ occur at $x=0$ or
$x=+\infty$, and (for uniform estimates) frequencies in their vicinity,
as well as the additional frequencies neglected here in case M.
The latter
should be treatable as in \cite{Er3}.
We expect that the problem of obtaining uniform estimates as $|\zeta|\to \infty$\footnote{Recall that we prove uniform estimates
for bounded $|\zeta|$ only.}
%may be recognized as the one-dimensional limit, which
can be treated
as a perturbation of the one-dimensional case using the arguments
of \cite{Z2}.
A remaining question is to obtain estimates that are uniform as $\zeta$ approaches the imaginary axis near Class III values.  At the moment we have such estimates  near Class I values on the imaginary axis.
We plan to address these questions in
a separate work.

Though they were not needed in order to obtain
useful instability results,
we regard the treatment of these remaining cases, and subsequent
applications to positive stability results,
as a very interesting open problem.
Another interesting open problem is to explore the implications
of these and the present
analyses for the associated viscous problem, as in \cite{Z1,TZ}.

\begin{rem}\label{o}
\textup{One can try to study the behavior near turning points using oscillatory integrals.  Near turning points of increasing or decreasing type these have the form associated with fold caustics and lead to Airy functions (see also Remark \ref{beh}). In future work we will explore whether such an approach might yield a simpler, or more efficient, or more unified treatment.}
\end{rem}

%%%%%%%%%%%%%%%%%%%%%%%%%%%%%%%%%%%%%%%%%%%%%%5

\section{Exact decaying solutions on $[a,\infty)$, $a\geq 0$}

\emph{\quad}In this section we prove Theorem \ref{conjugation} for the unbounded interval.   The theorem gives an accurate approximating formula when $\eps$ is large for the exact decaying solution $\theta(x,\zeta,\nu)$ of \eqref{a7} on the interval $[x_{*}+\delta,\infty)$ for Class III values of $\zeta$ and on the interval $[0,\infty)$ for all other values of $\zeta$ in $\Re\zeta\geq 0$.

\subsection{Construction of exact solutions}\label{exact}

\emph{\quad}Recall from \eqref{a9} that
\begin{align}\label{2ac}
\begin{split}
&\Phi_0(x,\zeta)=\{A_x^{-1}(x)\cdot (\zeta I+iA_y(x))\}^t\\
&\Phi_1(x,\nu)=\{A_x^{-1}(x)\cdot \nu I\}^t + \{ A_x^{-1}(x)B(x)\}^t,
\end{split}
\end{align}
where the coefficient matrices are defined in section \ref{coefficients}.
So with $h=\frac{1}{\eps}$ we can write
\begin{align}
\Phi_0(x,\zeta)+h\Phi_1(x,\nu)=\Phi_0(x,\zeta+h \nu)+h\{ A_x^{-1}(x)B(x)\}^t.
\end{align}
The eigenvalues of $\Phi_0(x,\zeta+h\nu)$ are $\mu_j(x,\zeta+h \nu)$, $j=1,\dots,5$, for $\mu_j$ as defined in \eqref{a10}.

In preparation for the theorem, we need to examine the effect of $h\{ A_x^{-1}(x)B(x)\}^t$ on the eigenvalues of $\Phi_0(x,\zeta)+h\Phi_1(x,\nu)$.
Direct computation and the use of  Assumptions \ref{profile} and \ref{rate} shows that\footnote{The computation is done on pages 114-117 of \cite{Er3}.  There is a  sign error in the expression for $\hat e_5\cdot W_{11}$,  which should be multiplied by $-1$, in (A.15) on p117 of \cite{Er3}.  This correction  yields the expression for row $5$ in \eqref{b1}.}
%On page 114 (extended) E writes $B=B_0+B_1$, where $B_0=O(e^{-\beta x})$.
%so $\eta\{ A_x^{-1}(x)B_0(x)\}^t$  is harmless from the point of view of our Lemma A.2.
%E computes $\{ A_x^{-1}(x)B_1(x)\}^t$ on pages 116,117 (extended).
%The rows of this matrix are given in (A.15), \emph{but note there is a
%  crucial sign error; the expression for $e_5\cdot W_{11}$ should be
%  multiplied by -1.}
\begin{align}\label{b1}
\{ A_x^{-1}(x)B(x)\}^t=O(e^{-\beta x})+\begin{pmatrix}0\\\mathrm{row\;5}\end{pmatrix}, \text{ where }\mathrm {row\; 5} = (*,*,*,*,-r_\lambda/u),
\end{align}
a matrix whose only nonzero row at equilibrium is the fifth.  This gives
\begin{align}\label{2add}
\Phi_0(x,\zeta)+h\Phi_1(x,\nu)=\Phi_0(x,\zeta+h \nu)+h\begin{pmatrix}0\\\mathrm{row\;5}\end{pmatrix}+O(h e^{-\beta x}).
\end{align}

From \eqref{phi0} we see that  $\Phi_0(x,\zeta)$  is a matrix of the form
\begin{align}\label{2ae}
\Phi_0=\begin{pmatrix}A&0\\B&\zeta/u\end{pmatrix}
\end{align}
where the eigenvalues of $A$ are $\mu_1$, $\mu_2$, $\mu_3=\mu_4=\zeta/u$.
Thus, the eigenvalues of $\Phi_0(x,\zeta)+h\Phi_1(x,\nu)$ are
\begin{align}\label{2ak}
\begin{split}
&\mu_j^*=\mu_j(x,\zeta+h\nu)+O(h e^{-\beta x}),\; j=1,2,3,4\\
&\mu_5^*=\mu_3(x,\zeta+h\nu)-h \frac{r_\lambda}{u}+O(h e^{-\beta x}), \text{ where }r_\lambda<0.
\end{split}
\end{align}
Here we are using \eqref{2add} and the fact that the  eigenvalues of $\Phi_0 +h\begin{pmatrix}0\\ \mathrm{row\;5}\end{pmatrix}$, which are semisimple, undergo a perturbation of size $O(h e^{-\beta x})$ in response to a matrix perturbation of the same size.

\begin{theo}\label{conjugation}
With $h=1/\eps$ consider the system \eqref{a7}
\begin{align}\label{6a}
\theta'=\frac{1}{h}\left[\Phi_0(x,\zeta)+h\Phi_1(x,\nu)\right]\theta
\end{align}
on an interval $[a,\infty)$, $a\geq 0$, and for values of $\zeta$,$\nu$ such that
\begin{align}\label{6b}
|\mu_1(x,\zeta+h\nu)-\mu_j(x,\zeta+h\nu)|\geq C_\zeta>0,\;j=2,\dots,5 \text{ for }0<h\leq h(\zeta,\nu) \text{ small enough}.
\end{align}
Then there exists an exact solution $\theta(x,\zeta,\nu,h)$ such that for any $\delta_*<\beta$
\begin{align}\label{6c}
\left|\theta-e^{\frac{1}{h}\int^{ x}_0 \mu_1^\sharp(s,\zeta,\nu,h)ds}\left[Te_1+O(h)\right]\right|\leq C_\zeta h e^{-\delta_* x} |e^{\frac{1}{h}\int^{x}_0 \mu_1^\sharp( s,\zeta,\nu,h)d s}|\;\;\text{ on }[a,\infty),
\end{align}
where $T=T(x,\zeta)$ is given by \eqref{a12}, $e_1=(1,0,0,0,0)$, and
\begin{align}\label{6aa}
\mu_1^\sharp=\mu_1(x,\zeta+h\nu)+O(he^{-\beta x}).
\end{align}
\end{theo}

%\begin{rem}
%1.  The hypothesis \eqref{6b} holds in all the cases described just before the Theorem. It fails for $\zeta$ of type III on %$[x_*,\infty)$, because $\mu_1(x_*,\zeta)=\mu_2(x_*,\zeta)$ in that case.

%2.  The result uses  Lemma \ref{vargaplem} and shows that we do not need to apply the Tracking Lemma in this paper.

%3.  The Theorem shows that we do not need to distinguish between $\zeta$ of type I and type II; in both cases we obtain an exact %decaying solution $\theta$ on $[0,\infty)$ just by applying Lemma \ref{vargaplem}.   There are only two kinds of $\zeta$ to %consider: $\zeta$ with $\Re\zeta=0$ for which there exist points $x_*$ such that $\zeta^2+c_0^2\eta(x_*)=0$, and all other %$\zeta$.
%\end{rem}

\textbf{Separation of eigenvalues. } Before giving the proof of Theorem \ref{conjugation} we check the hypothesis \eqref{6b} for all $\zeta$.  Since $|\nu|\leq R$ and we  take $h$ small, it suffices to check \eqref{6b} when $\nu=0$. We have
\begin{align}\label{differences}
\begin{split}
&\mu_2(x,\zeta)-\mu_1(x,\zeta)=\frac{2\kappa s}{\eta u}\\
&\mu_3(x,\zeta)-\mu_1(x,\zeta)=\frac{\zeta+\kappa s}{\eta u}.
\end{split}
\end{align}
Since $s(x,\zeta)=\sqrt{\zeta^2+c_0^2\eta}$ and
\begin{align}
\Re\sqrt{\zeta^2+c_0^2\eta}\geq \Re \zeta,
\end{align}
it follows in view of Assumption \ref{profile} that when $\Re \zeta>0$,   hypothesis \eqref{6b} is satisfied on $[0,\infty)$.
%We have $\Re \zeta>0$ for $\zeta$ of type II and $\zeta$ of type I for which $\Re\zeta>0$.
When  $\Re\zeta=0$ and $|\zeta|>\sup_x c_0\eta^{1/2}$,  we have
\begin{align}\label{b4}
|\zeta|>|s(x,\zeta)|=\sqrt{|\zeta|^2-c_0^2\eta}\geq C_\zeta >0,
\end{align}
so \eqref{6b} holds on $[0,\infty)$.  When $\Re\zeta=0$ and $|\zeta|< \inf_x c_0\eta^{1/2}$, then
\begin{align}\label{b5}
|\zeta|>s=\sqrt{c_0^2\eta-|\zeta|^2}\geq C_\zeta>0,
\end{align}
so again \eqref{6b} holds on $[0,\infty)$.

Finally, consider $\zeta\in \text{Class III}\setminus\{\pm ic_0\eta^{1/2}(\infty)\}$.   In cases M and D \eqref{b4} holds on $[x_*+\delta,\infty)$ (where $x_*$ denotes the rightmost turning point when there are two turning points corresponding to $\zeta$ in case M),
while in case I \eqref{b5} holds on $[x_*+\delta,\infty)$.   Thus, for $\zeta$ in Class III \eqref{6b} holds on $[x_*+\delta,\infty)$.

\begin{rem}\label{b5z}
1)\textup{ Observe that for $\zeta$ in a compact subset $K\subset \left(\{\Re\zeta\geq 0\}\setminus \text{Class III}\right)$ the constants $C_\zeta$ in \eqref{b4}, \eqref{b5} can be replaced by a uniform constant $C_K>0$. Thus, for $\zeta\in K$ and $|\nu|\leq R$, the separation inequality
\eqref{6b} holds with uniform constants $C_K$ and $h(K,R)$ on $[0,\infty)$.}

2) \textup{Similarly, one obtains uniform constants $C_K$, $h(K,R)$ in \eqref{6b} on $[x_*+\delta,\infty)$ for $\zeta$ in a compact set $K$ of the allowed  set of Class III values for each type of profile.}

\end{rem}

\begin{proof}[Proof of Theorem \ref{conjugation}]

\textbf{1. }We change to $\tilde x$ coordinates ($x=\tilde x  h$, $ h=\frac{1}{\eps}$), using $(\cdot)$ to denote $\frac{d}{d\tilde x}$ and $(')$ to denote $\frac{d}{dx}$. The spectral separation \eqref{6b} allows us to define a smooth conjugator
$S_1(\tilde x  h,\zeta,\nu,h)$ such that
\begin{align}\label{6d}
S_1^{-1}\left(\Phi_0(\tilde x h,\zeta)+ h\Phi_1(\tilde x h,\nu)\right)S_1=\begin{pmatrix}\mu_1^*&0\\0&G_1\end{pmatrix}:=\cG_1,
\end{align}
where $\mu_1^*$ is given in \eqref{2ak}, and the eigenvalues of $G_1$ are $\mu_j^*$, $j=2,\dots,5$ as in \eqref{2ak}.  We can take the first column of $S_1$ to satisfy
\begin{align}\label{6ee}
S_1e_1=Te_1+O(h).
\end{align}

Since the eigenvalues of $G_1$ satisfy
\begin{align}\label{6e}
\Re\mu_j^* \geq O(he^{-\beta h \tilde x})
\end{align}
we can find another smooth conjugator $S_2(\tilde x  h,\zeta,\nu,h)=\begin{pmatrix}1&0\\0&\cS_2\end{pmatrix}$ such that
\begin{align}\label{6f}
S_2^{-1}\cG_1S_2=\begin{pmatrix}\mu_1^*&0\\0&G_2\end{pmatrix}=\cG_2,
\end{align}
where
\begin{align}
\Re G_2\geq O(h e^{-\beta h\tilde x}).
\end{align}

    Set $S_3=S_1S_2$. Then  the function $W$ defined by
$\theta=S_3W$ satisfies
\begin{align}
\dot W=(\cG_2-h S_3^{-1}S_3') W.
\end{align}
Applying \eqref{6b} again,  we can find a conjugator of the form $S_4(\tilde x  h,\zeta,\nu,h)=I+h \cS_4$ such that $Y$ defined by $W=S_4Y$ satisfies
\begin{align}
\dot Y=\begin{pmatrix}\mu_1^\sharp&0\\0&G_3\end{pmatrix}-h^2 BY,
\end{align}
where $B=O(e^{-\beta \tilde x h})$ and
\begin{align}
\begin{split}
&\Re G_3=\Re G_2+O(h e^{-\beta \tilde x h})\geq O(h e^{-\beta h \tilde x})\\
&\mu_1^\sharp=\mu_1^*+O(h e^{-\beta \tilde x h}).
\end{split}
\end{align}

\textbf{2. }Next we set $Y=e^{\int^{\tilde x}_0 \mu_1^\sharp(\tilde s h,\zeta,\nu,h)d\tilde s}Z$ and observe that $Z$ satisfies

\begin{align}
\dot Z= \begin{pmatrix}0&0\\0&G_3-\mu_1^\sharp I\end{pmatrix}Z-h^2 BZ,
\end{align}
where
\begin{align}
\Re (G_3-\mu_1^\sharp I)\geq O(h e^{-\beta \tilde x h}).
\end{align}
%Now the result follows by applying Lemma \ref{vargaplem}, tracing back through the conjugations, and using \eqref{6ee}.
We can now apply Lemma \ref{vargaplem} (with $\delta_*$ any number less than $\beta$) to obtain a solution $Z$ satisfying
\begin{align}\label{2w}
|Z-e_1|\leq Ch e^{-\delta_*\tilde xh}
\end{align}
by \eqref{Pdecay2new}.  Tracing back we find
\begin{align}\label{2x}
|Y-e^{\int^{\tilde x}_0 \mu_1^\sharp(\tilde sh,\dots)d\tilde s}e_1|\leq Ch e^{-\delta_*\tilde xh} |e^{\int^{\tilde x}_0 \mu_1^\sharp(\tilde sh,\dots)d\tilde s}|,
\end{align}
so
\begin{align}\label{2y}
|W-e^{\int^{\tilde x}_0 \mu_1^\sharp(\tilde sh,\dots)d\tilde s}\left[e_1+h\cS_4 e_1\right]|\leq C_\zeta h e^{-\delta_*\tilde xh} |e^{\int^{\tilde x}_0 \mu_1^\sharp(\tilde sh,\dots)d\tilde s}|.
\end{align}
Since $\theta=S_3W$ and $S_3e_1=Te_1$,
\begin{align}\label{2z}
|\theta-e^{\int^{\tilde x}_0 \mu_1^\sharp(\tilde sh,\dots)d\tilde s}\left[Te_1+O(h)\right]|\leq C_\zeta h e^{-\delta_*\tilde xh} |e^{\int^{\tilde x}_0 \mu_1^\sharp(\tilde sh,\dots)d\tilde s}|\text{  on }[\frac{a}{h},\infty).
\end{align}
Switching back to $x$ coordinates gives (with $s=\tilde s h$)
\begin{align}\label{2aa}
|\theta-e^{\frac{1}{h}\int^{ x}_0 \mu_1^\sharp(s,\dots)ds}\left[Te_1+O(h)\right]|\leq C_\zeta h e^{-\delta_* x} |e^{\frac{1}{h}\int^{x}_0 \mu_1^\sharp( s,\dots)d s}|\;\;\text{ on }[a,\infty).
\end{align}

\end{proof}

As an immediate corollary of Theorem \ref{conjugation} we have
\begin{cor}\label{notIII}

When $\zeta\notin \text{Class III}$, the exact decaying solution $\theta(x,\tau,\eps)$ of \eqref{a7} satisfies
\begin{align}\label{2A}
|\theta(0,\tau,\eps)-T(0,\zeta)e_1|\leq C_\zeta/\eps \text{ for }\eps \geq \eps(\zeta) \text{ sufficiently large },
\end{align}
where $\tau=\eps\zeta+\nu$. Thus, the first column of the matrix \eqref{a12}, evaluated at $x=0$, gives the desired approximation.

For $\zeta$ in a compact subset $K\subset \left(\{\Re\zeta\geq 0\}\setminus \text{Class III}\right)$ and $|\nu|\leq R$ the constants $C_\zeta$, $\eps(\zeta)$ in \eqref{2A}  can be replaced by  uniform constants $C_K$, $\eps(K,R)$.

\end{cor}
\begin{proof}
For such $\zeta$ the formula of Theorem \ref{conjugation} applies on $[0,\infty)$. The uniformity statement follows
from part (1) of Remark \ref{b5z}, the uniform choice of $h_0$ in  Lemma \ref{vargaplem}, and inspection of the proof of Theorem \ref{conjugation}.
\end{proof}

\begin{rem}
\textup{Theorem \ref{conjugation} and its corollary do not require real-analyticity of the profile $w$, and are true without the restriction that $w$ be of type I, D, or M.}
\end{rem}

%\emph{Note that when E writes $\mu_1$ he always means $\mu_1(x,\zeta)$, never $\mu_1(x,\zeta+\frac{v}{\eps})$.}

\subsection{Comparison of exact and approximate solutions for $\zeta\in \text{Class III}$.}\label{compare}

\emph{\quad}When $\zeta$ lies in Class III, a matching argument is needed to analytically continue the exact solution $\theta$ on $[x_*+\delta,\infty)$, or an appropriate multiple thereof, around the turning point $x_*$.   To carry this out we must first examine how $\theta$ compares to the approximate solution $\theta_1=T\pi_1$ \eqref{a13} for $x\geq x_*+\delta$.
%We do the comparison for profiles of type D and I.  We will see that the result in case D can be applied also in case M.

The expression for $\theta_1$ is given in \eqref{a13}-\eqref{a15}
\begin{align}\label{2af}
\theta_1(x,\tau,\eps)=e^{\frac{1}{h}\int^{ x}_0 \mu_1(s,\zeta)ds+\int^x_{0,C_w}E_{11}(s,\zeta,\nu)ds}\left[T(x,\zeta)e_1+O(h)\right].
\end{align}
We assume $\zeta\in \text{Class III}_+$.
In case D (resp. I) we take $C_w=C_-$ (resp. $C_+$), a contour on the real axis except for a short excursion around $x_*$ in the lower (resp. upper) half plane.  In the $E_{11}$ integral for case M, a turning point where $d>0$  is excised via the upper half plane and one where $d<0$ via the lower half plane.\footnote{When there are two turning points $x_{2*}<x_{1*}$ in case M, starting with $s=i|s|$ in $x<x_{2*}$, analytic continuation along such a contour gives $s=|s|$ on $x_{2*}<x<x_{1*}$ and $s=i|s|$ on $x>x_{1*}$.}

\textbf{Cases D and M.}  In case D the formula \eqref{2aa} for $\theta$  holds on $[x_*+\delta,\infty)$ where
\begin{align}\label{3bb}
\mu_1^\sharp=\mu_1(x,\zeta+\nu h)+O(he^{-\beta x}),\;\beta>0.
\end{align}
With $\mu_1=\mu_1(x,\zeta)$ and $\mu_1^\sharp$ as in \eqref{3bb} here and in the rest of this section, we may write
\begin{align}\label{3bz}
e^{h^{-1}\int^x_0\mu_1^\sharp}Te_1=e^{h^{-1}\int^x_0\mu_1+\int^x_{0,C_-} E_{11}+h^{-1}\int^x_0(\mu_1^\sharp-\mu_1)-\int^x_0 E_{11}}Te_1=e^{H_\eps(x,\zeta,\nu)}\theta_1(1+O(h))
\end{align}
where
\begin{align}\label{3bbb}
H_\eps(x,\zeta,\nu):=\eps\int^x_0(\mu_1^\sharp-\mu_1)-\int^x_{0,C_-} E_{11}.
\end{align}
Thus, \eqref{2aa} and \eqref{3bz} imply
\begin{align}\label{3dd}
|\theta - e^{H_\eps(x,\zeta,\nu)}\theta_1|\leq \frac{C}{\eps}|e^{H_\eps(x,\zeta,\nu)}\theta_1| \text{ for }x\geq x_*+\delta.
\end{align}

The inequality \eqref{3dd} holds also in case M, where $x_*$ now denotes the rightmost turning point in cases where there are two turning points.  We do not treat the special value $\zeta=ic_0\eta^{1/2}(x_M)$, where $x_M$ is the location of the maximum.

Next we examine the growth rate of $e^{H_\eps}$.
\begin{prop}\label{3dz}
For profiles of type D and $\zeta\in \text{Class III}_+\setminus\{c_0\eta^{1/2}(\infty)\}$, the factor  $e^{H_\eps(x,\zeta,\nu)}$ satisfies the estimate:
\begin{align}\label{3d}
C_1 e^{\Re(\alpha_1 \frac{i\nu^2}{\eps}) x}\leq |e^{H_\eps(x,\zeta,\nu)}|\leq C_2 e^{\Re(\alpha_2 \frac{i\nu^2}{\eps}) x}\text{ for }x\in [0,\infty),\;x\neq x_*,
\end{align}
for some positive constants $C_i$, $\alpha_i$ depending on $x_*$ and the profile $w$, but which can be taken independent of $\eps$, and $\nu$, and a compact subset of $\zeta$ as above.   For $\Re \nu>0$ and  $\Im v>0$ (resp. $\Im v<0$) \eqref{3d}  represents exponential decay (resp. growth) in $x$.

For profiles of type M, we also exclude the special value $\zeta=ic_0\eta^{1/2}(x_M)$. Then the  estimate \eqref{3d} holds away from turning points with uniform constants for a compact set of $\zeta$.

\end{prop}

\begin{proof}
\textbf{1. }
Since $\Re\sqrt{(\zeta+h\nu)^2+c_0^2\eta}\geq \Re (\zeta+h\nu)$ we have
\begin{align}\label{3b}
\begin{split}
&\Re\mu_1(x,\zeta+h\nu)=\Re \frac{-\kappa\left(\kappa(\zeta+h\nu)+\sqrt{(\zeta+h\nu)^2+c_0^2\eta}\right)}
{\eta u}\leq\\
&\qquad\qquad\quad\qquad-\frac{\kappa^2}{\eta u}\Re h\nu-\frac{\kappa}{\eta u}\Re (\zeta+h\nu)=-\left(\frac{\kappa^2+\kappa}{\eta u}\right)\Re h\nu.
\end{split}
\end{align}
For $x\leq x_*+\delta$, $x\neq x_*$,  it follows from \eqref{3b} and the formula for $E_{11}$ \eqref{E11}
that there exist  positive constants $C_i$ as described such that
\begin{align}\label{3by}
C_1 \leq |e^{H_\eps(x,\zeta,\nu)}|\leq C_2.
\end{align}

\textbf{2. }Now assume $x\geq x_*+\delta$.  For such $x$ we write
\begin{align}\label{3bx}
\begin{split}
&H_\eps(x,\zeta,\nu)=H_\eps(x_*+\delta,\zeta,\nu)+H_\eps^\flat(x,\zeta,\nu)\text{ where }\\
&H_\eps^\flat(x,\zeta,\nu):=\frac{1}{h}\int^x_{x_*+\delta}(\mu_1^\sharp-\mu_1)-\int^x_{x_*+\delta} E_{11}.
\end{split}
\end{align}

Using \eqref{3b}  and $\Re \mu_1(x,\zeta)=0$ (since $w$ is of type D), we see that
\begin{align}\label{3c}
\Re\left(\frac{1}{h}(\mu_1^\sharp-\mu_1)- E_{11}\right)= -\frac{\kappa^2}{\eta u}\Re \nu-\frac{\kappa}{\eta u}\frac{1}{h}\Re\sqrt{(\zeta+h\nu)^2+c_0^2\eta}-\Re E_{11}+O(e^{-\beta x}).
\end{align}
Assumptions \ref{profile}, \ref{rate} and the formula for $E_{11}$ \eqref{E11} imply (with $s=\sqrt{\zeta^2+c_0^2\eta}=i|s|$)
\begin{align}\label{3cw}
 E_{11}(x,\zeta,\nu)=-\frac{\kappa^2}{\eta u} \nu-\frac{\kappa}{\eta u}\frac{\zeta}{s}\nu+ O(e^{-\beta x}).
\end{align}
The function $s$ is bounded away from $0$ for $x\geq x_*+\delta$, so we can
expand the square root in \eqref{3c} about $\zeta^2+c_0^2\eta=s^2$ obtaining:
\begin{align}\label{3h}
\frac{1}{h} \sqrt{(\zeta+h\nu)^2+c_0^2\eta}=\frac{1}{h} s(1+\frac{h\nu\zeta}{ s^2}+\frac{h^2\nu^2}{2 s^2}-...).
\end{align}
Since $s=i|s|$, we get after noting cancellations
\begin{align}\label{3hz}
\Re\left(\frac{1}{h}(\mu_1^\sharp-\mu_1)- E_{11}\right)= \frac{\kappa}{2\eta u|s|}\Re(ih\nu^2)+ O(h^2)+O(e^{-\beta x})
\end{align}
for $x\geq x_*+\delta$.  In view of \eqref{3by}, \eqref{3bx}, and Assumption \ref{profile}, this implies the estimate \eqref{3d} with possibly new $\C_i$.

\textbf{3. } With $\zeta=ic_0\eta^{1/2}(x_M)$ excluded and $H_\eps$ redefined as described before the Theorem, the proof for case M is the same as above.

\end{proof}

\begin{rem}\label{o3}
\textup{For rate functions with $r_v\neq 0$ at equilibrium and satisfying the weaker assumption discussed in Remark \ref{rate3}, we see from the formula for $E_{11}$ that there will be an extra non-decaying term in \eqref{3cw} depending on $r_v$.  However, in \eqref{3hz} that term will be cancelled by a corresponding extra term in the expansion of $\mu_1^\sharp$.}

\end{rem}

\textbf{Case I.} In this case we have $s(x,\zeta)=\Re s\geq C>0$ for $x\geq x_*+\delta$,  and computations almost identical to those above show that
\begin{align}
\Re\left(\frac{1}{h}(\mu_1^\sharp-\mu_1)- E_{11}\right)= \frac{-\kappa}{2\eta u s}\Re(h\nu^2)+O(h^2)+O(e^{-\beta x})
\end{align}
for $x\geq x_*+\delta$.
\begin{prop}\label{3dx}
For profiles of type I and $\zeta\in \text{Class III}\setminus\{c_0\eta^{1/2}(\infty)\}$, the factor  $e^{H_\eps(x,\zeta,v)}$ satisfies the estimate:
\begin{align}\label{3hx}
C_1 e^{\Re(-\alpha_1 \frac{\nu^2}{\eps}) x}\leq |e^{H_\eps(x,\zeta,\nu)}|\leq C_2 e^{\Re(-\alpha_2 \frac{\nu^2}{\eps}) x}\text{ for }x\in [0,\infty),\;x\neq x_*,
\end{align}
for some positive constants $C_i$, $\alpha_i$ depending on $x_*$ and the profile $w$, but which can be taken independent of $(\nu,\eps)$ and a compact subset of $\zeta$ as above.   For $\Re \nu>0$ and  $|\Im \nu|<\Re \nu$ (resp. $|\Im \nu|>\Re\nu\geq 0$) \eqref{3hx} represents exponential decay (resp. growth) in $x$.
\end{prop}

%This discrepancy between $\theta$ and $\theta_1$ indicates that in subsequent matching arguments we will need to replace $\theta$ %by a suitable multiple $M(\eps,\zeta,v)$ of it.

\begin{rem}\label{3f}
 \textup{In cases I, D, and M the estimate \eqref{3dd} implies that the statement $M(\eps,\zeta,v)\theta\sim\theta_1$  (``$\sim$"  as defined below \eqref{a20}) is not true for any multiple $M(\eps,\zeta,\nu)$ of $\theta$ on $[x_*+\delta,b)$ for any $b\in (x*+\delta,\infty]$.}
\end{rem}

\section{Exact solutions on bounded regions near a turning point}

\emph{\quad}This section is devoted to proving Theorem \ref{mpp}, which provides a sufficient condition for approximate solutions $\theta_i$ of $\theta'=-P^t\theta$, defined by formulas like \eqref{a13},  to be close to exact solutions
when $\eps$ is large.
Theorem \ref{mpp} was informally stated  in \cite{Er3} with a reference to a set of 1954 N.Y.U. Notes by K. O. Friedrichs.  The result is referred to in \cite{Er3} as the ``method of the parameter problem".  The proof we give here is an adaptation to ODEs on bounded regions of the complex plane of a similar argument in Chapter 6 of \cite{CL} for ODEs on bounded intervals of the real line.

\subsection{Method of the parameter problem.}

\emph{\quad} Consider a general $N\times N$ system with parameter $\eps>>1$\footnote{Here we are for convenience suppressing
dependence of $\Phi$ on additional parameters in the notation. All the arguments below work when there is continuous dependence on a compact set of parameters.}
\begin{align}\label{5a}
\theta'=\eps\Phi(z,\eps)\theta
%\text{ where }\Phi=\Phi_0+\frac{\Phi_1}{\eps}
\end{align}
on some bounded, simply connected region $\cH\subset\CC$.  For $i=1,\dots,N$
%CHANGED: need nonneg! Else, not only estimate tells nothing, but
%contraction argument is false... -KZ 1-29-2011
%and $m$ fixed, let
and $m\ge 0$ fixed, let
%ENDCHANGED
\begin{align}\label{5aa}
\theta_i=e^{\eps h_i(z)+k_i(z)}[b_{i0}+\eps^{-1}b_{i1}+\dots+\eps^{-(m+1)}b_{i(m+1)}]:=e^{q_i(z,\eps)}P_i(z,\eps)
\end{align}
and define $N\times N$ matrices
\begin{align}
Q=\diag(q_1,\dots,q_N),\;\; P=\{P_1,\dots,P_N\}.
\end{align}
We assume that $P^{-1}(z,\eps)$ is uniformly bounded for $z\in\cH$, $\eps$ large and that the $\theta_i$ are approximate solutions of \eqref{5a} in $\cH$  of order $m$ in the sense that\footnote{Thus, $\Phi_m=\frac{1}{\eps}(P'+PQ')P^{-1}$.}
\begin{align}\label{5b}
(Pe^Q)'=(P'+PQ')e^Q:=\eps\Phi_m Pe^Q= \eps\Phi Pe^Q +\eps (\Phi_m-\Phi)Pe^Q,
\end{align}
where
%CHANGED: corrected logical order.-KZ
\begin{align}\label{5cc}
|\Phi-\Phi_m|\leq \frac{C_1}{\eps^{m+2}}\text{ for }z\in \cH,
\end{align}
and thus  also
$|\eps (\Phi-\Phi_m)Pe^Q|\leq C\frac{|e^Q|}{\eps^{m+1}}$
for $z\in \cH$.
%%%%OLD
%\begin{align}
%|\eps (\Phi-\Phi_m)Pe^Q|\leq C\frac{|e^Q|}{\eps^{m+1}} \text{ for }z\in \cH.
%\end{align}
%Thus, the matrix $\Phi_m$ defined by the second equality in \eqref{5b} satisfies
%\begin{align}\label{5cc}
%|\Phi-\Phi_m|\leq \frac{C_1}{\eps^{m+2}}\text{ for }z\in \cH.
%\end{align}

%For convenience in exposition we now take $N=3$.
The following Theorem gives conditions under which we can find an exact solution $\theta$ of \eqref{5a} close to $\theta_1$ in the sense that for $\eps$ large and $z\in\cH$:
\begin{align}\label{5c}
|\theta-\theta_1|\leq C_m\frac{|e^{q_1(z,\eps)}|}{\eps^{m+1}}\text{ in }\cH.
\end{align}
%CHANGED: added this obvious statement-KZ
Of course the labeling of $\theta_j$ is arbitrary, hence the result
applies equally to $\theta_j$, $j\ne 1$.
%ENDCHANGED

\begin{theo}\label{mpp}
Consider the $N\times N$ system \eqref{5a} with  parameter $\eps>>1$ on the bounded, simply connected region $\cH\subset\bC$.  Let $\cH'$ be a bounded, simply connected neighborhood of $\cH$, and suppose that  $\Phi$ is analytic for $z\in\cH'$. Suppose that the approximate solutions $\theta_i$ given by \eqref{5aa} are analytic in $\cH'$  and
%CHANGED: same thing... -KZ
%satisfy \eqref{5cc},
satisfy \eqref{5cc} for some $m\ge 0$,
%ENDCHANGED
and that $P^{-1}(z,\eps)$ is uniformly bounded for $z\in\cH'$ and $\eps$ large.

a.) For each $j\in \{2,\dots,N\}$ suppose there exists a point $z_j\in \cH'$ such that \emph{any} point $z\in \cH$ can be joined to $z_j$ by a path lying in $\cH'$ on which $\Re (h_j-h_1)(\tau,\eps)$ decreases (not necessarily strictly) as $\tau$ moves from $z_j$ towards $z$.    For any $z\in \cH$ let $\cP_j(z)$ be such a path starting from $z_j$ and suppose
\begin{align}\label{5dd}
\sum_{j=2}^N \;\sup_{z\in\cH}|\cP_j(z)|\leq L<\infty,
\end{align}
where $|\cP_j(z)|$ denotes the length of $\cP_j(z)$.
Then there exists an exact solution $ \theta$ of \eqref{5a} in $\cH$  such that \eqref{5c} holds.

b.) There is an obvious analogue of part (a) (not requiring analyticity) for the case where $\cH$ is taken to be a bounded closed interval
%CHANGED: added, to ref. below-KZ
$[a,b]$
%ENDCHANGED
on the real axis and the paths $\cP_j$ are chosen inside that interval.
%CHANGED: addded-
In this case existence of admissible paths is equivalent to the
assumption that for any fixed $j\ne 1$, either $\Re \mu_j\le \Re \mu_1$ or else
$\Re \mu_j\ge \Re \mu_1$ on all of $[a,b]$,
i.e., there is a neutral spectral gap between $\mu_1$ and
all other $\mu_j$.
\end{theo}

\begin{proof}
For convenience we take $N=3$ in the proof.
Pick $z_2$ and $z_3$ satisfying the assumptions in Theorem \ref{mpp} and define $3\times 3$ matrices
\begin{align}\label{5d}
P^{(2)}=[P_1,P_2,0],\;\;\;P^{(3)}=[0,0,P_3]
\end{align}
satisfying
\begin{align}
P=P^{(2)}+P^{(3)}.
\end{align}
Observe that $\theta_1$ is a solution of $\theta'=\eps\Phi_m\theta$ and that \eqref{5a} may be written as
\begin{align}\label{5e}
\theta'=\eps\Phi_m\theta+\eps(\Phi-\Phi_m)\theta.
\end{align}

We construct a solution of \eqref{5e} in $\cH$  as a solution $\theta(z,\eps)$ of the integral equation
\begin{align}\label{5f}
\begin{split}
&\theta(z,\eps)=\theta_1(z,\eps)+\eps P^{(2)}(z,\eps)e^{Q(z,\eps)}\int^z_{z_2}e^{-Q(\tau,\eps)}\left[P^{-1}(\tau,\eps)\left(\Phi(\tau,\eps)-\Phi_m(\tau,\eps)\right)\right]\theta(\tau,\eps)d\tau+\\
&\quad\quad \eps P^{(3)}(z,\eps)e^{Q(z,\eps)}\int^z_{z_3}e^{-Q(\tau,\eps)}\left[P^{-1}(\tau,\eps)\left(\Phi(\tau,\eps)-\Phi_m(\tau,\eps)\right)\right]\theta(\tau,\eps)d\tau,
\end{split}
\end{align}
where the first and second integrals are on paths $\cP_2(z)$, $\cP_3(z)$, respectively.
Note that the only exponentials appearing in the elements of $P^{(2)}(z,\eps)e^{Q(z,\eps)-Q(\tau,\eps)}$ are $e^{q_j(z,\eps)-q_j(\tau,\eps)}$, $j=1,2$, and the only exponential appearing in the elements of $P^{(3)}(z,\eps)e^{Q(z,\eps)-Q(\tau,\eps)}$ is $e^{q_3(z,\eps)-q_3(\tau,\eps)}$.

We solve \eqref{5f} by the iteration scheme
\begin{align}\label{5g}
\begin{split}
&\theta_{(l+1)}(z,\eps)=\theta_1(z,\eps)+\eps P^{(2)}(z,\eps)e^{Q(z,\eps)}\int^z_{z_2}e^{-Q(\tau,\eps)}\left[P^{-1}(\tau,\eps)\left(\Phi(\tau,\eps)-\Phi_m(\tau,\eps)\right)\right]\theta_{(l)}(\tau,\eps)d\tau+\\
&\quad\quad \eps P^{(3)}(z,\eps)e^{Q(z,\eps)}\int^z_{z_3}e^{-Q(\tau,\eps)}\left[P^{-1}(\tau,\eps)\left(\Phi(\tau,\eps)-\Phi_m(\tau,\eps)\right)\right]\theta_{(l)}(\tau,\eps)d\tau,
\end{split}
\end{align}
where $\theta_{(0)}=0$.    We claim
\begin{align}\label{5h}
|(\theta_{(l+1)}-\theta_{(l)})e^{-q_1(z,\eps)}|\leq C_2 \; \;\frac{L}{\eps^{m+1}}\;\sup_{\tau\in\cH}|(\theta_{(l)}-\theta_{(l-1)})e^{-q_1(\tau,\eps)}|,
\end{align}
%CHANGED: added for clarity-KZ
%$m\ge 0$,
$m\ge 0$,
%ENDCHANGED
for $C_2$ depending on the norms of $P$,  $P^{-1}$, and $C_1$ (from \eqref{5cc}), as well as the $\Re k_i$.  In estimating the contribution to the right side of \eqref{5h} coming from the first integral in \eqref{5g}, for example, we have used
\eqref{5cc}, \eqref{5dd}, the uniform boundedness of $P^{-1}$, and the fact that
\begin{align}\label{5hz}
|e^{q_2(z,\eps)-q_2(\tau,\eps)-q_1(z,\eps)+q_1(\tau,\eps)}|=e^{\Re\left(q_2(z,\eps)-q_2(\tau,\eps)-q_1(z,\eps)+q_1(\tau,\eps)\right)}\leq M_2\text{ for }\tau\in\cP_2(z),
\end{align}
where $M_2$ depends only on $\Re k_i$, $i=1,2$.
Thus, for $\eps$ large we obtain a contraction and the $\theta_{(l)}$ converge uniformly on $\cH$ to a solution of the integral equation.   Moreover, if $|\theta_1e^{-q_1}|\leq C_0$, then for $\eps$ large enough \eqref{5h} implies
\begin{align}\label{ti}
|\theta_{(l)}e^{-q_1(z,\eps)}|\leq 2C_0.
\end{align}
From the integral equation we then obtain
\begin{align}
|(\theta-\theta_1) e^{-q_1(z,\eps)}|\leq  C_2 \; \;\frac{2C_0 L}{\eps^{m+1}}.
\end{align}

\end{proof}

\begin{defn}\label{admissible}
a) Let $\theta_i$ be an approximate solution of \eqref{5a} satisfying \eqref{5b} for some $m\geq 0$ in a region $\cH\subset\bC$.  We say that $\theta_i$ is an \emph{approximate solution of order $m$} in $\cH$.

b)  Suppose that $\theta_i$ as above has been shown to be close to an exact solution on some subregion $\cH_a\subset\cH$ in the sense of \eqref{5c}.  Then we say that $\theta_i$ is \emph{admissible} in $\cH_a$.   The curves defining the boundary of the admissible subregion $\cH_a$ are often referred to as \emph{Stokes curves}.
\end{defn}

We will see below that the admissible region $\cH_a$ is sometimes a proper subregion of $\cH$.  For example, in applying Theorem \ref{mpp} to justify an approximate solution $\theta_1$ in the case where $N=3$, it may happen that one can choose paths $\cP_2(z)$ satisfying the requirements of the Theorem for all $z\in\cH$, but that one can choose paths $\cP_3(z)$ only for $z$ in a proper subregion $\cH_a$ of $\cH$.  In that case $\theta_1$ is admissible only in $\cH_a$.

\begin{defn}\label{paths}

When applying Theorem \ref{mpp} to test the admissibility of $\theta_1$, we refer to paths on which $\Re(h_j-h_1)$ decreases as \emph{$\cP_j$ paths} for $j\in\{2,\dots,N\}$.  More generally, when testing the admissibility of $\theta_k$, we refer to paths on which $\Re(h_j-h_k)$ decreases as \emph{$\cP_j$ paths} for $j\in \{1,\dots,N\}\setminus \{k\}$.  The context (i.e., choice of $\theta_k$ being tested) should prevent confusion.
\end{defn}

\begin{rem}\label{samepaths}
\textup{Consider again the $5\times 5$ system
\begin{align}\label{5rb}
\theta'=-P^t(x,\tau,\eps)=[\eps\Phi_0(x,\zeta)+\Phi_1(x,\nu)]\theta
\end{align}
with $\mu_j$, $j=1,\dots,5$ as in \eqref{a10}.  When applying Theorem \ref{mpp} to test the admissibility of $\theta_1$, for example, observe that since $\mu_3=\mu_4=\mu_5$, the $\cP_3$ paths, $\cP_4$ paths, and $\cP_5$ paths can all be taken to be the same.  In this situation we usually refer to all of these paths as $\cP_3$ paths.}

\end{rem}

\subsection{Analytic continuation and the choice of integration paths.}\label{choice}

\emph{\quad}  We now fix $\zeta=i|\zeta|\in \text{Class III}$.
By Assumptions \ref{thermo} and \ref{profile} the coefficients in the system \eqref{5rb} extend analytically to a neighborhood of $[0,\infty)$. Quantities appearing in the definition of $\theta_i$ \eqref{a13} are of two types: those that depend real-analytically on the profile $w$, and those like $\mu_1$, $\mu_2$, and certain terms in $T(x,\zeta)$ \eqref{a12} and $E_{11}$ \eqref{E11} that depend on $s(x,\zeta)=\sqrt{\zeta^2+c_0^2\eta(x)}$.   Terms of the first type extend analytically to a neighborhood of $[0,\infty)$, while analytic extensions of terms of the second type have branch points at turning points where $s(x_*,\zeta)=0$.

 When testing the admissibility of a solution $\theta_i$ in a region $\cH$ whose points all lie close to a turning point $x_*$ where $c_0^2\eta$ is either increasing or decreasing,  there is a simple procedure given in \cite{Er3} for choosing paths that satisfy the requirements of Theorem \ref{mpp}.  We need to use this procedure in section \ref{match}, so we illustrate it here for the reader's convenience in the increasing case where
\begin{align}\label{5rc}
d:=\frac{d(c_0^2\eta)}{dx}(x_*)>0.
\end{align}

We take $\cH$ to be an open disk centered at $x_*$ from which a small (closed) neighborhood of the segment $[0,x_*]$ has been removed, and explain first how to choose $\cP_2$ paths and $\cP_3$ paths for showing the admissibility of $\theta_1$. Suppose that $\theta_1$ is defined before continuation into $\cH$ by formula \eqref{2af} for $x>x_*$.

 Expanding $s^2(z)=\zeta^2+c_0^2\eta(z)$ about $x_*$, we compute from the formulas \eqref{differences}:
\begin{align}\label{5rd}
\begin{split}
&\mu_2(z)-\mu_1(z)=(2\kappa d^{\frac{1}{2}}/\eta u)(x_*)(z-x_*)^{\frac{1}{2}}[1+O(z-x_*)]\\
&\mu_3(z)-\mu_1(z)=(i/\eta^{\frac{1}{2}}\kappa)(x_*)[1+O((z-x_*)^{\frac{1}{2}}],
\end{split}
\end{align}
where $(z-x_*)^{\frac{1}{2}}$ denotes the branch that is positive for $z>x_*$.   We set
\begin{align}\label{5re}
h_{ji}=h_j-h_i, \;\;h_{ji}^*=h_{ji}(x_*),
\end{align}
and integrate \eqref{5rd} to obtain
\begin{align}\label{5rf}
\begin{split}
&h_{21}(z)-h_{21}^*=(4\kappa d^{\frac{1}{2}}/3\eta u)(x_*)(z-x_*)^{\frac{3}{2}}+O((z-x)^{\frac{5}{2}})\\
&h_{31}(z)-h_{31}^*=(i/\eta^{\frac{1}{2}}\kappa)(x_*)(z-x_*)+O((z-x_*)^{\frac{3}{2}}).
\end{split}
\end{align}

Before choosing paths on which $\Re(h_{j1}-h_{j1}^*)$ (and therefore also $\Re h_{j1}$) is nonincreasing, we first locate the \emph{transition rays} where $\Re(h_{j1}-h_{j1}^*)\approx 0$, $j=2,3$.  With $\phi=\arg(z-x_*)$ these rays are
\begin{align}\label{5rg}
\phi_{21}=\pm\frac{\pi}{3}, \;\pi;\;\;\;\phi_{31}=0,\pi,
\end{align}
where the subscripts have the obvious meaning. The transition rays defined by $\phi_{21}$ determine $3$ sectors centered at $x_*$:
\begin{align}\label{5rh}
\cS_0:-\frac{\pi}{3}<\phi<\frac{\pi}{3},\quad \;\cS_1: \frac{\pi}{3}<\phi<\pi,\quad\;\cS_{-1}:-\pi<\phi<-\frac{\pi}{3}.
\end{align}

Paths in $\cH$ on which $\Re(h_{21}-h_{21}^*)$ is nonincreasing are easy to determine if one knows the level curves of $\Re(h_{21}-h_{21}^*)$.  In sector $\cS_0$, $\Re(h_{21}-h_{21}^*)$ is positive with level curves that are roughly parallel to the union of the two rays $\phi=\frac{\pi}{3}$ and $\phi=-\frac{\pi}{3}$. In sectors $\cS_{\pm 1}$ $\Re(h_{21}-h_{21}^*)$ is negative and the level curves have a similar description.   $\cP_2$ paths with initial point $z_2>x_*$ can now be constructed by piecing together segments that lie on level curves of $\Re(h_{21}-h_{21}^*)$ with (correctly oriented) segments that are transverse to level curves.  If $\cH$ is small enough,  any point $z\in \cH$ can be reached by such a path starting from $z_2$.

From \eqref{5rf} and \eqref{5rg} we see that the level curves of $\Re(h_{31}-h_{31}^*)$ are well-approximated by horizontal lines near $x_*$ and
$\Re(h_{31}-h_{31}^*)>0$ in $\Im(z-x_*) <0$.
Thus, if we choose $z_3$ near $x_*$ in $\Im(z-x_*)  <0$ and if $\cH$ is small enough, any point $z\in\cH$ can be reached by
a $\cP_3$ path starting at $z_3$.

A similar procedure applies when testing the admissibility of approximate solutions $\theta_j$, $j=2,\dots,5$.  For example, when testing $\theta_2$ in $\cH$ one chooses $\cP_1$ paths and $\cP_3$ paths using\footnote{The choice of branch of $(z-x_*)^\frac{1}{2}$ in these formulas depends both on the region $\cH$ and on the specification of $\theta_2$ before its analytic continuation into $\cH$. Typically, $\theta_j$ is specified first on either $x<x_*$ or $x>x_*$.}
\begin{align}\label{5ri}
\begin{split}
&h_{12}(z)-h_{12}^*=-(4\kappa d^{\frac{1}{2}}/3\eta u)(x_*)(z-x_*)^{\frac{3}{2}}+O((z-x_*)^{\frac{5}{2}})\\
&h_{32}(z)-h_{32}^*=(i/\eta^{\frac{1}{2}}\kappa)(x_*)(z-x_*)+O((z-x_*)^{\frac{3}{2}}).
\end{split}
\end{align}

\section{Matching arguments for $\zeta\in\text{Class III}$}\label{match}

 \emph{\quad}In Theorem \ref{conjugation} we constructed for $\zeta\in\text{Class III}$ an exact decaying solution $\theta$ of equation \eqref{6a} on $[x_*+\delta,\infty)$.  In this section we give the matching arguments needed to determine  the analytic continuation of that solution up to $x=0$.
 %Recall that in the matching arguments, we must exclude the Class III values in \eqref{a18b}.
%This is accomplished by matching arguments, which (by a symmetry property of the Evans function)  are needed only for %$\zeta=i|\zeta|$ of type III.
%\begin{align}\label{5qy}
%\{\pm i \lim_{x\to \infty}c_0\eta^{\frac{1}{2}}.
%\end{align}

\subsection{Case I: $c_0^2-u^2$ is strictly increasing. }\label{matchI}
\emph{\quad}In order to carry out the matching argument we must first use Theorem \ref{mpp} to construct a basis of exact solutions on bounded domains near $x_*$.

Let $\theta_i(x,\zeta,\nu,\eps)$, $i=1,\dots,5$,  be approximate solutions constructed in the standard way and defined before analytic continuation for $x>x_*$ by formulas
\begin{align}\label{5q}
\theta_i(x,\zeta,\nu,\eps)=\exp{[\eps h_i(x,\zeta)+k_i(x,\zeta,\nu)]}a_i(x,\zeta,\nu),
\end{align}
where
\begin{align}\label{5r}
\begin{split}
&h_i(x,\zeta)=\int^x_{0}\mu_i(\tau,\zeta)d\tau, \text{ for all }i,\\
&k_i(x,\zeta,\nu)=\int^x_{0,C_+} E_{ii}(\tau,\zeta,\nu)d\tau, \text{ for } i =1,2,
\end{split}
\end{align}
and the subscript $C_+$ indicates a contour on the real axis except for a small excursion in the upper half plane to avoid $x_*$
In \eqref{5r} the square root $s(x,\zeta)=i|s|$ for $x<x_*$, $s=|s|$ for $x>x_*$, and is given by analytic continuation on $C_+$.
The functions $E_{11}$ and $E_{22}$ have terms with $(z-x_*)^{-j}$  singularities at $x_*$, where $j\in\{1/2,1\}$.
On the other hand $\mu_i$ is bounded near $x_*$.
%so $\int^x_0\mu_i=\int^x_{0,C_+}\mu_i$.  We omit the other formulas for quantities associated to $\theta_3$, ..., $\theta_{n+4}$, %(which correspond to the multiple eigenvalue $\mu_3$), because we will not need them.

\begin{defn}
For $\delta>0$ small and any set $S\subset \bC$, we define a $\delta-$neighborhood of $S$ to be the union of open balls of radius $\delta$ centered at points of $S$.
\end{defn}

Below we let $N_\phi$ denote the closure of a $\delta-$neighborhood of the ray $\{z:\arg (z-x_*)=\phi\}$.
Let $B\subset \bC$ denote a $4\delta$-neighborhood of the segment  $[x_*-6\delta,x_*]$.\footnote{For the matching arguments to be valid we need to work near $x_*$, so the parameter $\delta$ introduced here is to be fixed sufficiently small.  For the continuation of solutions all the way to $x=0$ see step \text{4} of the proof of Proposition \ref{88}.}

Let $\theta_{1,M}$ be an approximate solution defined by analytically continuing $\theta_1$ to $B\setminus N_\pi$, and   denote by $\bar\theta_{1,M}$  the corresponding exact solution justified by Theorem
\ref{mpp} in this region.   Paths of type $\cP_2$ in the sense of Theorem \ref{mpp} can be chosen starting at some $z_2>x_*$, while $\cP_3$ paths can be chosen starting just below $x_*$ in $\Im z<0$.   Here one can use exactly the same paths as those constructed in section \ref{choice} using \eqref{5rf}.

Denote by $\bar\theta_2$ an exact solution constructed by a similar procedure, but based on the approximate solution $\theta_2$ analytically continued to $B\setminus N_{\pi/3}$.
Paths of type $\cP_1$ can be chosen starting at a point $z_1$ in the 3rd quadrant (relative to $x_*$), while $\cP_3$ paths can be chosen starting in $\Im z<0$.

Similarly, we let $\bar\theta_3,\bar\theta_4,\bar\theta_{5}$ be exact solutions associated to analytic continuations of $\theta_3,\theta_4,\theta_{5}$ and defined in
$B\setminus N_{-\pi/3}$.  In this case paths of types $\cP_1$ and $\cP_2$ can be chosen to start directly above $x_*$ on the axis $\Re z= x_*$.

Let $\theta_1^+$ denote  an approximate solution obtained by analytically continuing $\theta_1$ to $B\setminus N_{-\pi/3}$ and let
$\theta_1^-$ denote an an approximate solution obtained by analytically continuing $\theta_1$ to $B\setminus N_{\pi/3}$.
Each of these solutions can be justified by Theorem \ref{mpp} on a  $2\delta$-neighborhood $\cO_{2\delta}$  of the segment $B\cap [0,x_*-3\delta]$, yielding corresponding exact solutions
$\bar\theta_1^+$ and $\bar\theta_1^-$.  For $\bar\theta_1^+$ the paths of type $\cP_2$ and $\cP_3$ start in the third quadrant, while for $\bar\theta_1^-$ paths of type $\cP_2$ start in the second quadrant and $\cP_3$ paths start in the third quadrant.

It is helpful for computing $V(\tau,\eps)$ to have explicit formulas for $\theta_1^\pm(z,\zeta,\nu,\eps)$, $z\in\cO_{2\delta}$.  These have the form
\begin{align}\label{5s}
\theta_1^\pm(z,\zeta,\nu,\eps)= \exp{[\eps h_1^\pm(z,\zeta)+k_1^\pm(z,\zeta,\nu)]}a_1^\pm(z,\zeta,\nu)
\end{align}
where
\begin{align}\label{5t}
\eps h_1^\pm(z,\zeta)+k_1^\pm(z,\zeta,\nu):=\int^z_{0,\cP_\pm}[\eps\mu_1(\tau,\zeta)+E_{11}(\tau,\zeta,\nu)]d\tau
\end{align}
and $a_1^\pm$ are obtained from $T(x,\zeta)e_1$, $x>x_*$, by analytic continuation.
The path $\cP_+$ consists of the segment $[0,x_*-2\delta]$ followed by any path in $\cO_{2\delta}$ from  $x_*-2\delta$  to $z$, and the function of $z$ given by $s(z,\zeta)=\sqrt{\zeta^2+c_0^2\eta(z)}$ appearing in $\mu_1$ is
obtained from the branch taking positive values on $x>x_*$ by analytic continuation in $B_+\setminus N_{-\pi/3}$.  The same square root, call it $s_+(z,\zeta)$ is used, of course, wherever it appears in $a_1^+$.   It is readily checked that in $\cO_{2\delta}$ the function $\theta_1^+$ defined by \eqref{5s} equals $\theta_1^+$ as originally defined.

The path $\cP_-=\cP_{-,a}\cup\cP_{-,b}$, where $\cP_{-,a}$ is a segment from $0$ to $x_*-2\delta$, followed by a semicircle in $\Im z>0$ terminating at $x_*+2\delta$.  The path $\cP_{-,b}$ starts at $x_*+2\delta$, continues on the semicircle in $\Im z<0$ terminating at $x_*-2\delta$, and continues from  $x_*-2\delta$ along any path in
$\cO_{2\delta}$ ending at $z$.  In the part of the integral \eqref{5t} on $\cP_{-,a}$ the square root $s_+(z,\zeta)$ is used, while on
$\cP_{-,b}$, one uses $s_-(z,\zeta)$, which is obtained from the branch taking positive values on $x>x_*$ by analytic continuation in $B_+\setminus N_{\pi/3}$.  The square root $s_-(z,\zeta)$ is used wherever $s$ appears in $a_1^-$.  In $\cO_{2\delta}$ the function $\theta_1^-$ defined by \eqref{5s} equals $\theta_1^-$ as originally defined.

Finally, observe that since $\mu_3(x,\zeta)=\frac{\zeta}{u}$, the function $h_3(x,\zeta)$ extends analytically to a neighborhood of $[0,\infty)$ containing $B$.
\begin{rem}\label{5tz}
\textup{The approximate solutions $\theta_1^\pm$ defined here in $B\setminus N_{\mp\frac{\pi}{3}}$ extend analytically (as approximate solutions) to a neighborhood of the segment $[0,x_*-2\delta]$, and explicit formulas for those extensions are given by the formulas \eqref{5s},\eqref{5t} using the obvious prolongations of the paths $\cP_\pm$.
The admissibility of the extensions on $[0,x_*-2\delta]$ is shown in step \textbf{4} of the proof of Proposition \ref{88}.}
\end{rem}

\begin{prop}\label{88}
For a profile of type I, fix $\zeta\in\text{Class III}_+\setminus\{ic_0\eta^{1/2}(0+), ic_0\eta^{1/2}(\infty)\}$.
Let $\theta$ be the exact decaying solution on $x\geq x_*+\delta$  constructed for case I in Theorem \ref{conjugation}, extended analytically to $[0,\infty)$.
For an appropriately selected multiple $\bar\theta_{1,G}=G(\eps,\zeta,\nu)\theta$ we have
on $[0,x_*-3\delta]$
\begin{align}\label{4q}
\bar\theta_{1,G}=\bar\theta_1^++\bar\theta_1^-+O(1/\eps).
\end{align}
Here $\bar\theta_1^+$ and $\bar\theta_1^-$ are analytic extensions of the exact solutions $\bar\theta_1^\pm$ defined above on $\cO_{2\delta}$ which satisfy
\begin{align}\label{444p}
|\bar\theta_1^\pm-\theta_1^\pm|\leq \frac{C_\zeta}{\eps}|\theta_1^\pm|\text{ on }[0,x_*-3\delta],
\end{align}
where $\theta_1^\pm$ are given by the formulas \eqref{5s},\eqref{5t}.

The constant $C_\zeta$ in \eqref{444p} can be replaced by $C_K$ for $\zeta$ in a compact subset $K$ of the allowed values.

\end{prop}

\begin{proof}
 \textbf{1. }
Let $x_1=x_*+2\delta$ and recall \eqref{3dd}
\begin{align}\label{4a}
|\theta - e^{H_\eps(x,\zeta,\nu)}\theta_1|\leq \frac{C}{\eps}|e^{H_\eps(x,\zeta,\nu)}\theta_1|, \text{ where }\theta_1=e^{\eps\int^x_0\mu_1+\int^x_{0,C_+} E_{11}} a_1.
\end{align}
Multiply through by
\begin{align}\label{4b}
G(\eps,\zeta,\nu):=e^{-H_\eps(x_1,\zeta,\nu)}
\end{align}
to get
\begin{align}\label{4c}
|G(\eps,\zeta,\nu)\theta - e^{H_\eps(x,\zeta,\nu)-H_\eps(x_1,\zeta,\nu)}\theta_1|\leq \frac{C}{\eps}|e^{H_\eps(x,\zeta,\nu)-H_\eps(x_1,\zeta,\nu)}\theta_1|,
\end{align}
Evaluating this at $x_1$ we obtain
\begin{align}\label{4d}
|G(\eps,\zeta,\nu)\theta - \theta_1|(x_1)\leq \frac{C}{\eps}|\theta_1|(x_1),
\end{align}

\textbf{2. }Next we carry out  the matching argument using
\begin{align}\label{4e}
\bar\theta_{1,G}(x,\zeta,\nu,\eps):=G(\eps,\zeta,\nu)\theta
\end{align}
for the exact decaying solution on $[x_1,\infty)$.  Observe that the exact solutions $\bar\theta_{1,G}$ and
$\bar\theta_{1,M},\bar\theta_2,\dots,\bar\theta_{5}$, have original domains (i.e., before analytic continuation) that include
$[x_*+2\delta,x_*+4\delta)$.\footnote{All of these exact solutions can be analytically continued as solutions to a neighborhood of $[0,\infty)$, but it is only on the original domains that we have information about asymptotic behavior as $\eps\to\infty$.  The original domain of $\bar\theta_{1,M}$, for example, is $B\setminus N_\pi$.}
Using the basis $\{\bar\theta_{1,M},\bar\theta_2,\dots,\bar\theta_{5}\}$ we have
\begin{align}\label{4f}
\bar\theta_{1,G}=c_1\bar\theta_{1,M}+c_2\bar\theta_2+\dots+c_{5}\bar\theta_{5},
\end{align}
where for all $i$, $c_i=c_i(\eps,\zeta,\nu)$. By Theorem \ref{mpp} we have
\begin{align}\label{4g}
|\bar\theta_{1,M}-\theta_1| \leq \frac{C}{\eps}|\theta_1| \text{ on }[x_*+2\delta,x_*+4\delta)
\end{align}
and analogous statements hold for the other $\bar\theta_i$.   An estimate like \eqref{4g} is not true for $\bar \theta_{1,G}$, but at $x_1$ we have from \eqref{4d}
\begin{align}\label{4h}
|\bar \theta_{1,G}-\theta_1|(x_1)\leq \frac{C}{\eps}|\theta_1|(x_1).
\end{align}

The coefficients in \eqref{4f} are independent of $x$ so we can evaluate equation \eqref{4f} at $x_1$ to determine them.   Recall $\Re h_3(x_1)=0$ and let
\begin{align}\label{4hh}
a:=-\Re h_1(x_1)>0,\;b=\Re h_2(x_1)>0.
\end{align}
A direct analysis\footnote{This statement about the coefficients is proved by writing $\bar\theta_{1,G}=e^{\eps h_1+k_1}a_1+O(e^{\eps h_1+k_1}/\eps)$ and similar expressions for the other solutions, plugging into equation \eqref{4f},  using the linear independence of $a_1,\dots,a_{5}$, and taking account of the growth rates of the exponentials.  The straightforward details are omitted.} of the coefficients in \eqref{4f} using \eqref{4g} and \eqref{4h} shows that
\begin{align}\label{4i}
c_1=1+O(\eps^{-1}), \;\;c_2=O(e^{-(a+b)\eps}/\eps),\;\; c_3=O(e^{-a\eps}/\eps),\;\;\dots,\;\;c_{5}=O(e^{-a\eps}/\eps).
\end{align}
%So
%\begin{align}\label{4j}
%\bar\theta_{1,G}=(1+O(\eps^{-1}))\;\bar\theta_{1,M}+O(e^{-(a+b)\eps})\bar\theta_2+O(e^{-a\eps})\bar\theta_3+\dots+O(e^{-a\eps})\bar\theta_{n+4}.
%\end{align}
The  functions of $x$ in \eqref{4f} extend analytically to a neighborhood of $[0,\infty)$, and so \eqref{4f} holds on that neighborhood.

\textbf{3. }From the earlier construction we already know the asymptotic behavior as $\eps\to \infty$ of $\bar\theta_2,\dots,\bar\theta_{5}$ for $B\cap \{x<x_*\}$, but that is not the case for $\bar\theta_{1,M}$, whose behavior we know only on
$B\setminus N_\pi$.   To remedy this we expand $\bar\theta_{1,M}$ in terms of the basis
$\{\bar\theta_1^+, \bar\theta_1^-,\bar\theta_3,\dots,\bar\theta_{5}\}$ on the intersection of the original domains:
\begin{align}\label{4jj}
\cI:=(B\setminus N_\pi)\;\cap\;\cO_{2\delta}\;\cap\;(B\setminus N_{-\pi/3}).
\end{align}
This gives
\begin{align}\label{4k}
\bar\theta_{1,M}=c_+\bar\theta_1^++c_-\bar\theta_1^-+d_3\bar\theta_3+\dots+d_{5}\bar\theta_{5},
\end{align}
where again all the coefficients depend on $(\eps,\zeta,\nu)$. Next we evaluate equation \eqref{4k} at a point $\uz\in\cI\cap \{\Im z<0\}$, and use \eqref{5t}, \eqref{5rf}, and $h_{31}^*=0$ to see that
\begin{align}\label{4l}
a_1^+:=\Re h_1^+(\uz)<0< a_1^-:=\Re h_1^-(\uz)<a_3:=\Re h_3(\uz).
\end{align}
We have
\begin{align}\label{4m}
C_1e^{a_1^+\eps}\leq |\bar\theta_1^+(\uz,\zeta,\nu,\eps)|\leq C_2e^{a_1^+\eps},
\end{align}
and estimates analogous to \eqref{4m} at $\uz$ hold for $\{\bar\theta_{1,M},\bar\theta_1^-\}$ and $\{\bar\theta_3,\bar\theta_4,\bar\theta_{5}\}$ with $a_1^+$ replaced by $a_1^-$ and $a_3$ respectively.

Using
\begin{align}\label{4mz}
|\bar\theta_{1,M}-\theta_1^-|(\uz)\leq \frac{C}{\eps}|\theta_1^-|(\uz)
\end{align}
and an analysis of coefficients similar to the one that gave \eqref{4i}, we find
\begin{align}\label{4n}
c_-=1+O(1/\eps),\;|c_+|\leq C e^{(a_1^--a_1^+)\eps},\;d_j=O(e^{(a_1^--a_3)\eps}), j=3,4,5.
\end{align}

To get more information on $c_+$ we select  $z^\sharp\in \cI\cap\{\Im z>0\}$, evaluate equation \eqref{4k} at $z^\sharp$, and define
\begin{align}\label{4o}
b_1^-:=\Re h_1^-(z^\sharp)<0< b_1^+:=\Re h_1^+(z^\sharp), \text{ and }b_3:=\Re h_3(z^\sharp)<0.
\end{align}
Now $\{\bar\theta_{1,M},\bar\theta_1^+\}$ grow like $e^{b_1^+\eps}$ at $z^\sharp$ (in the sense of \eqref{4m}), $\bar\theta_1^-$ grows like $e^{b_1^-\eps}$, and $\{\bar\theta_3,\bar\theta_4,\bar\theta_{5}\}$ grow like  $e^{b_3\eps}$.
Thus, when equation \eqref{4k} is evaluated at $z^\sharp$, all terms to the right of $c_+\bar\theta_1^+$ decay exponentially fast in $\eps$.  So we conclude from
\begin{align}\label{4my}
|\bar\theta_{1,M}-\theta_1^+|(z^\sharp)\leq \frac{C}{\eps}|\theta_1^+|(z^\sharp)
\end{align}
%the known asymptotic behavior of $\bar\theta_{1,M}$ and $\bar\theta_1^+$ at $z^\sharp$
that
$c_+=1+O(\frac{1}{\eps})$.   Substituting \eqref{4k} into \eqref{4f} we obtain
\begin{align}\label{4oo}
\bar\theta_{1,G}=(1+O(1/\eps))\;\bar\theta_1^++(1+O(1/\eps))\;\bar\theta_1^-+k_2\bar\theta_2+\dots+k_{5}\bar\theta_{5},
\end{align}
where the $k_j$ decay exponentially fast in $\eps$.

\textbf{4. }
At this point we know the asymptotic behavior of the functions on the right in \eqref{4oo} only near $x_*$.  For example we know
\begin{align}\label{4p}
|\bar\theta_1^\pm-\theta_1^\pm|\leq \frac{C}{\eps}|\theta_1^\pm|
\end{align}
only in $\cO_{2\delta}$, a $2\delta-$neighborhood of the segment $B\cap [0,x_*-3\delta]=(x_*-10\delta,x_*-3\delta]$.
To conclude that \eqref{4p} holds on all of $[0,x_*-3\delta]$ we observe that $\bar\theta_1^\pm$ can be constructed by Theorem \ref{mpp} on
$[0,x_*-5\delta]$ (say) using $\cP_2$ and $\cP_3$ paths approaching zero along the $x-$axis, but which  start at the \emph{same} initial points already chosen for the justification on $\cO_{2\delta}$.   This works because $\Re\mu_j=0$ on $[0,x_*-5\delta]$ for $j=1,2,3$.
The other functions on the right in \eqref{4oo} are treated similarly.

\textbf{5. } The ability to choose $C_\zeta$ uniformly for $\zeta$ in a compact subset of the allowed $\text{Class III}$ values follows from the corresponding uniformity statements in part (2) of Remark \ref{b5z} and the comparison proposition Prop. \ref{3dx}, and inspection of the proof of Theorem \ref{mpp} together with steps \textbf{1-4} above.

\end{proof}

\textbf{A simpler formula for $\theta_1^+$ and $\theta_1^-$ on $[0,x_*-3\delta]$}.   It follows directly from the definition of $\theta_1^+$ given earlier that
\begin{align}\label{4r}
\theta_1^+(x,\zeta,\nu,\eps)=\exp\left[\int^x_0\left(\eps\mu_1(\tau,\zeta)+E_{11}(\tau,\zeta,\nu)\right)d\tau\right]a_1(x,\zeta,\nu)\text{ on }[0,x_*-3\delta],
\end{align}
where $s(x,\zeta)=i|s|$ whenever it occurs in \eqref{4r}.  Let us define
\begin{align}\label{4s}
\theta_2^+(x,\zeta,\nu,\eps)=\exp\left[\int^x_0\left(\eps\mu_2(\tau,\zeta)+E_{22}(\tau,\zeta,\nu)\right)d\tau\right]a_2(x,\zeta,\nu)\text{ on }[0,x_*-3\delta]
\end{align}
using the same $s(x,\zeta)$; here $a_j=T(x,\zeta)e_j$, $j=1,2$.  Observe that by changing the sign of $s$ in the formula for $\mu_1$ we obtain $\mu_2$; similarly, we obtain $a_2$ from $a_1$ and $E_{22}$ from $E_{11}$ by changing the sign of $s$ (see section \ref{coefficients}).
It  follows immediately from this and the formulas for $\theta_1^-$ given in \eqref{5s}, \eqref{5t} that
\begin{align}\label{4t}
\theta_1^-=\alpha(\eps,\zeta,\nu)\theta_2^+\text{ on }[0,x_*-3\delta].
\end{align}
Here the factor $\alpha$ is given by
\begin{align}\label{4u}
\alpha(\eps,\zeta,\nu):=\exp\left[\int_C(\eps \mu_1+E_{11})d\tau\right],
\end{align}
where $C$ proceeds from $0$ along the $x-$axis to $x_*-2\delta$, then around $x_*$ on a clockwise circle of radius $2\delta$, then back to $0$ along the $x-$axis.
The square root changes by analytic continuation along the contour, starting on $[0,x_*-2\delta]$ with $s=i|s|$ as above\footnote{On the return trip from $x_*-2\delta$ to $0$ we  have $s=-i|s|$}.

%An easy application of part (b) of Theorem \ref{mpp} yields an exact solution $\bar\theta_2^+$  such that
%\begin{align}\label{4uu}
%|\bar\theta_2^+-\theta_2^+|\leq \frac{C}{\eps}|\theta_2^+|\text{ on }[0,x_*-3\delta].
%\end{align}

Defining the exact solution $\bar\theta_2^+:=\alpha^{-1}\bar\theta_1^-$, we have:
\begin{cor}\label{explicit}
For a profile of type I, fix $\zeta\in\text{Class III}_+\setminus\{ ic_0\eta^{1/2}(0+), ic_0\eta^{1/2}(\infty)\}$.
On $[0,x_*-3\delta]$
\begin{align}\label{4v}
\bar\theta_{1,G}(x,\zeta,\nu,\eps)=\bar\theta_1^++\alpha(\eps,\zeta,\nu)\bar\theta_2^++O(1/\eps),
\end{align}
where $\alpha$ is given by \eqref{4u} and
\begin{align}\label{4vz}
|\bar\theta_j^+-\theta_j^+|\leq \frac{C}{\eps}|\theta_j^+|,\;\;j=1,2
\end{align}
on $[0,x_*-3\delta]$. In particular, we have
\begin{align}\label{4vr}
|\bar\theta_{1,G}(0,\zeta,\nu,\eps)-\left[T(0,\zeta)e_1+\alpha(\eps,\zeta,\nu)T(0,\zeta)e_2\right]|\leq C_\zeta/\eps
 \end{align}
 for $T$ as in \eqref{a12}.  The constant $C_\zeta$ can be replaced by $C_K$ for $\zeta$ in a compact subset $K$ of the allowed values.
\end{cor}

\begin{rem}\label{size}
\textup{Since $\mu_i$, $i=1,2$ are purely imaginary on $[0,x_*-3\delta]$, the functions $\theta^+_j$, $j=1,2$ are $O(1)$ there as $\eps\to\infty$. Thus, the error term in \eqref{4v} is small compared to the other terms.}
\end{rem}

\subsection{Case D: $c_0^2-u^2$ is strictly decreasing. }\label{matchD}

\emph{\quad} To construct the exact solutions needed for the matching argument in this case we start with approximate solutions $\theta_i$ given by
\begin{align}\label{7a}
\theta_i(x,\zeta,\nu,\eps)=\exp{[\eps h_i(x,\zeta)+k_i(x,\zeta,\nu)]}a_i(x,\zeta,\nu),
\end{align}
but now they are defined before analytic continuation for $x<x_*$ where
\begin{align}\label{7b}
\begin{split}
&h_i(x,\zeta)=\int^x_{0}\mu_i(\tau,\zeta)d\tau, \text{ for all }i\\
&k_i(x,\zeta,\nu)=\int^x_{0} E_{ii}(\tau,\zeta,\nu)d\tau, \text{ for } i =1,2,
\end{split}
\end{align}
and $s(x,\zeta)=|s|$ for $x<x_*$.\footnote{The value of $s$ in $x>x_*$, $s=i|s|$, is obtained from that in $x<x_*$ by continuation in $\Im z<0$.}

Again let $B\subset \bC$ denote a $4\delta$-neighborhood of the segment  $[x_*-6\delta,x_*]$, but now let $ \tilde N_{ \phi}$ denote a closed $\delta-$neighborhood of the ray $\{z:\arg (x_*-z)=\phi\}$.
Let $\theta_{1,M}$ be an approximate solution defined by analytically continuing $\theta_1$ to $B\setminus \tilde N_{-\pi/3}$, and   denote by $\bar\theta_{1,M}$  the corresponding exact solution justified by Theorem
\ref{mpp} in that region.\footnote{Caution: The ray   $\arg (x_*-z)=-\pi/3$, for example, lies in the second quadrant relative to $x_*$.}
When choosing $\cP_2$ paths, for example, we now use\footnote{Formula \eqref{7bz} corrects a sign error in \cite{Er3}, p.49.}
\begin{align}\label{7bz}
h_{21}(z)-h_{21}^*=-(4\kappa (-d)^{\frac{1}{2}}/3\eta u)(x_*)(x_*-z)^{\frac{3}{2}}+O((x_*-z)^{\frac{5}{2}})
\end{align}
in place of \eqref{5ri}.
Paths of type $\cP_2$ and $\cP_3$ can be chosen starting in the fourth quadrant (relative to $x_*$).

Let $\theta_2$ be an be an approximate solution defined by analytically continuing $\theta_2$ to $B\setminus \tilde N_{-\pi/3}$.  Paths of type $\cP_3$ terminating at any point in this region can be chosen with a common initial point in the fourth quadrant.
Paths of type $\cP_1$  terminating at any point in the region $\cB_1:=\{z:\frac{\pi}{3}<\arg (x_*-z)<\frac{5\pi}{3}\}$ can be chosen with a common initial point in the first quadrant.
Paths of type $\cP_1$  terminating at any point in the region $\cB_2:=\{z:-\frac{\pi}{3}<\arg (x_*-z)<\pi\}$ can be chosen with a common initial point on $x<x_*$.  By Theorem \ref{mpp} we obtain one exact solution $\bar\theta_{21}$ in $\cB_1$ and another $\bar\theta_{22}$  in $\cB_2$ satisfying
\begin{align}\label{5riz}
|\bar\theta_{2j}-\theta_2|\leq \frac{C}{\eps}|\theta_2|\text{ in }\cB_j, \;j=1,2.
\end{align}
Similarly, we let $\bar\theta_3,\bar\theta_4,\bar\theta_{5}$ be exact solutions associated to analytic continuations of $\theta_3,\theta_4,\theta_{5}$ and justified in
$B\setminus \tilde N_{\pi/3}$.  In each case paths of types $\cP_1$ and $\cP_2$ can be chosen to start directly above $x_*$ on the axis $\Re z= x_*$.

Now the goal is to prove
\begin{prop}\label{44l}
For a profile of type D, fix $\zeta\in\text{Class III}_+\setminus\{ ic_0\eta^{1/2}(0+), ic_0\eta^{1/2}(\infty)\}$.
Let $\theta$ be the exact decaying solution on $x\geq x_*+\delta$  constructed for case D in Theorem \ref{conjugation}, extended analytically to $[0,\infty)$.
For an appropriately selected multiple $\bar\theta_{1,G}=G(\eps,\zeta,\nu)\theta$ we have
on $[0,x_*-3\delta]$
\begin{align}\label{8a}
|\bar\theta_{1,G}-\theta_1|\leq \frac{C_\zeta}{\eps}|\theta_1|\text{ on }[0,x_*-3\delta].
\end{align}

The constant $C_\zeta$ in \eqref{444p} can be replaced by $C_K$ for $\zeta$ in a compact subset $K$ of the allowed values.

\end{prop}

\begin{proof}
\textbf{1. }As in case I we take $x_1=x_*+2\delta$, define
\begin{align}\label{8c}
G(\eps,\zeta,v)=e^{-H_\eps(x_1,\zeta,\nu)},
\end{align}
and note that $\bar\theta_{1,G}:=G(\eps,\zeta,\nu)\theta$ satisfies
\begin{align}\label{8d}
|\bar\theta_{1,G} - \theta_{1,M}|(x_1)\leq \frac{C}{\eps}|\theta_{1,M}|(x_1).
\end{align}

\textbf{2. }Using the basis $\{\bar\theta_{1,M},\bar\theta_{21},\bar\theta_3,\bar\theta_4,\bar\theta_{5}\}$ we have
\begin{align}\label{8e}
\bar\theta_{1,G}=c_1\bar\theta_{1,M}+c_2\bar\theta_{21}+\dots+c_{5}\bar\theta_{5}\text{ on }B\cap\{x\geq x_1\},
\end{align}
where for all $i$, $c_i=c_i(\eps,\zeta,v)$. By the construction of $\bar\theta_{1,M}$ we have
\begin{align}\label{8f}
|\bar\theta_{1,M}-\theta_{1,M}| \leq \frac{C}{\eps}|\theta_{1,M}| \text{ on }[x_*+2\delta,x_*+4\delta)
\end{align}
and analogous statements hold for the other functions on the right.

\textbf{3. } We have an integral formula
\begin{align}\label{8ff}
\theta_{1,M}(z,\zeta,\nu,\eps)= \exp{[\eps h_{1,M}(z,\zeta)+k_{1,M}(z,\zeta,\nu)]}a_{1,M}(z,\zeta,\nu)
\end{align}
where
\begin{align}\label{8g}
\eps h_{1,M}(z,\zeta):=\int^z_{0,\Gamma}\eps \mu_1(\tau,\zeta)d\tau,
\end{align}
 $\Gamma$ consists of the segment $[0,x_*-2\delta]$ followed by  any path in $B\setminus\tilde N_{-\pi/3}$ from $x_*-2\delta$ to $z$, and the definitions of $\mu_1(\tau,\zeta)$, $k_{1,M}(z,\zeta,v)$, and $a_{1,M}(z,\zeta,v)$
are readily gleaned from the definition of $\theta_{1,M}$ given above. We shall also need the corresponding formulas for
$\theta_{2}$ in $\cB_1$ and $\cB_2$ and the functions $\theta_3,\dots,\theta_{n+4}$ in $B\setminus\tilde N_{\pi/3}$.

\textbf{4. }We determine the behavior of the coefficients in \eqref{8e} after evaluating \eqref{8e} at $x_1$. From the properties of
the $\mu_i$ on the real axis and the integral formulas for the approximate solutions, we easily determine that
\begin{align}\label{8h}
a:=-\Re h_{1,M}(x_1,\zeta)>0,\;b:=\Re h_2(x_1,\zeta)>0, \text{ and }\Re h_3(x_1,\zeta)=0.
\end{align}
Thus we  have a situation like \eqref{4hh} and the same coefficient analysis shows
\begin{align}\label{8i}
c_1=1+O(\eps^{-1}), \;\;c_2=O(e^{-(a+b)\eps}/\eps),\;\; c_j=O(e^{-a\eps}/\eps),\;j=3,4,5.
\end{align}

\textbf{5. }We know the asymptotic behavior in $B\cap\{x\leq x_*-3\delta\}$ of all functions on the right in \eqref{8e} except for $\bar\theta_{21}$.  Thus we write
\begin{align}\label{8j}
\bar\theta_{21}=d_1\bar\theta_{1,M}+d_2\bar\theta_{22}+d_3\bar\theta_3+d_4\bar\theta_4+d_{5}\bar\theta_{5},
\end{align}
where again all coefficients depend on $(\eps,\zeta,v)$. Evaluate equation \eqref{8j} at a point $\uz$ given by
\begin{align}\label{8k}
\arg (x_*-\uz)=\frac{\pi}{3}+\delta,\; |\uz|=2\delta,
\end{align}
a point where all functions in \eqref{8j} are known asymptotically.  The integral formulas for the approximate solutions yield (with paths of integration mostly restricted to the $x-$axis)
\begin{align}\label{8l}
\tilde a:=-\Re h_{1,M}(\uz,\zeta)>0,\;\tilde b:=\Re h_2(\uz,\zeta)>0, \text{ and }\tilde c =\Re h_3(\uz,\zeta)\text{ with }|\tilde c|<c(\delta),
\end{align}
where $c(\delta)\to 0$ as $\delta\to 0$.
Direct analysis of the coefficients in \eqref{8j} using \eqref{5riz} gives
\begin{align}\label{8m}
|d_1|\leq C e^{(\tilde b+\tilde a)\eps}/\eps,\; d_2=1+O(1/\eps),\;|d_j|\leq C e^{(\tilde b-\tilde c)\eps}/\eps,\; j\geq 3.
\end{align}
For $\delta$ small we have $\tilde a<a$, $\tilde b<b$, so  from \eqref{8e}, \eqref{8i}, \eqref{8j}, and \eqref{8m} we conclude
\begin{align}\label{8n}
|\bar\theta_{1,G}-\bar\theta_{1,M}|\leq \frac{C}{\eps}|\bar\theta_{1,M}|\text{ on }B\cap [0,x_*-3\delta].
\end{align}

\textbf{6. }It remains to show that \eqref{8n} holds on $[0,x_*-3\delta]$.  The problem is that, while the functions on the right in
\eqref{8e} and \eqref{8j} have analytic extensions to $x=0$, we do not know the asymptotic  behavior in $\eps$ of those extensions.  As in case I we can use Theorem \ref{mpp} to
determine this behavior in the case of $\bar\theta_{1,M}$,  because $\Re(h_2-h_1)\text{ and }\Re(h_3-h_1)$ decrease as $x\to 0$ on $[0,x_*-3\delta]$. This allows us to choose
$\cP_2$ paths and $\cP_3$ paths terminating at points on $[0,x_*-3\delta]$  and starting at the same initial points already chosen for the earlier justification of $\bar\theta_{1,M}$ on
$B\cap \tilde (\tilde N \setminus -\frac{\pi}{3})$.    Thus we obtain
\begin{align}\label{8b}
|\bar\theta_{1,M}-\theta_1|\leq \frac{C}{\eps}|\theta_1|\text{ on }[0,x_*-3\delta].
\end{align}
This method does not work for the other functions appearing in \eqref{8e} and \eqref{8j}. For example, as $x$ decreases to $0$,
$\Re(h_1-h_2)$ and $\Re(h_3-h_2)$ both increase, and therefore Theorem \ref{mpp} cannot be used to show the admissibility of $\bar\theta_{22}$ on $[0,x_*-3\delta]$.

Instead we use the differential equation  to estimate directly how much these functions can grow as $x$ approaches $0$ in $[0,x_*-3\delta]$.  We then use the estimates on the coefficients in
\eqref{8i} and \eqref{8m} to show that in spite of this growth, \eqref{8n} still holds on $[0,x_*-3\delta]$.

We illustrate this argument in the case of $\bar\theta_{3}$.  On $[0,x_*-3\delta]$ we have for some constants $c$,$C$:
\begin{align}\label{8o}
\Re\mu_1\leq c<0=\Re\mu_3<C\leq \Re \mu_2.
\end{align}
After diagonalizing the system with the transformation $T(x,\zeta)$ \eqref{a12} and setting $\bar\theta_{3}=TW$, we obtain with $D=\diag(\mu_1,\dots,\mu_5)$:
\begin{align}\label{8oz}
d_x|W|^2=d_x(W,W)=2\Re(\eps DW,W)+O(|W|^2)\Rightarrow |W|_x\geq \eps \Re \mu_1 |W|-C^\sharp|W|.
\end{align}
So for $\eps$ large, $W$ and therefore $\bar\theta_{3}$ grows at most by a factor
\begin{align}\label{8p}
 e^{\eps\int^{x_*-3\delta}_y\left(|\Re\mu_1|(s,\zeta)+\frac{C^\sharp}{\eps}\right)ds}
\end{align}
as $x$ varies from  $x_*-3\delta$ to $y$, where $y\in[0,x_*-3\delta]$.   In the expansion of $\bar\theta_{1,G}$ on $B\cap [0,x_*-3\delta]$ the function $\bar\theta_3$ occurs with a coefficient of
$c_2d_3+c_3$.    Since $c_3=O(e^{-a\eps}/\eps)$ and
\begin{align}
a>\int^{x_*-3\delta}_0|\Re\mu_1|(s,\zeta)ds
\end{align}
we obtain
\begin{align}\label{8pz}
|c_3\bar\theta_3|(y)\leq \frac{C}{\eps}e^{-a\eps}e^{\eps\int^y_{x_*-3\delta}\Re \mu_1(s,\zeta)ds}\leq \frac{C}{\eps}e^{\eps\int^y_{0}\Re \mu_1(s,\zeta)ds}\leq \frac{C}{\eps}|\bar\theta_{1,M}|(y).
\end{align}
A similar estimate for $c_2d_3\bar\theta_3$ shows that the contribution of $\bar\theta_3$ on $[0,x_*-3\delta]$ is negligible.
The remaining functions in the expansion of $\bar\theta_{1,G}$ are treated similarly.  Thus we obtain
\begin{align}\label{8q}
|\bar\theta_{1,G}-\bar\theta_{1,M}|\leq \frac{C}{\eps}|\bar\theta_{1,M}|\text{ on } [0,x_*-3\delta],
\end{align}
which with \eqref{8b} implies the result.

\textbf{7. }The final uniformity statement is proved as in step \textbf{5} of the proof of Proposition \ref{88}.

\end{proof}

\begin{cor}\label{8qz}
Fix $\zeta\in\text{Class III}_+\setminus\{ic_0\eta^{1/2}(0+), ic_0\eta^{1/2}(\infty)\}$.  In case D we have
\begin{align}\label{8qy}
|\bar\theta_{1,G}(0,\zeta,\nu,\eps)-T(0,\zeta)e_1|\leq C_\zeta/\eps
\end{align}
for $T$ as in \eqref{a12}.   The constant $C_\zeta$ can be replaced by $C_K$ for $\zeta$ in a compact subset $K$ of the  specified Class III values.
\end{cor}

\subsection{Case M: $c_0^2-u^2$ has a single interior maximum.}

\emph{\quad} The nonvanishing of $d=\frac{d(c_0^2\eta)}{dx}(x_*)$ played an important role in the matching arguments for cases I and D.
In case M denote by $x_M$ the interior point where
\begin{align}
\frac{d(c_0^2\eta)}{dx}(x_M)=0.
\end{align}
A different type of matching argument is needed for the Class III value
\begin{align}\label{9a}
\zeta_M=ic_0\eta^{1/2}(x_M).
\end{align}
The turning point problem for $\zeta_M$ is studied in \cite{Er3}, section 6, pages 71-98,  but we shall not treat it here.  For all other non-exceptional Class III values of $\zeta$ in case M, the turning point problems that arise can be handled using the earlier results for cases I and D, together with their proofs.

For each such $\zeta$ there are either one ($x_*=x_{1*}$) or two ($x_{2*}<x_{1*}$) turning points .  As in cases I and D we define $\bar\theta_{1,G}=G(\eps,\zeta,\nu)\theta$, where $\theta$ is the decaying solution constructed in Theorem \ref{conjugation} on $[x_{1*}+\delta,\infty)$ and
\begin{align}\label{9b}
G(\eps,\zeta,\nu)=e^{-H_\eps(x_1,\zeta,\nu)}, \;x_1=x_{1*}+\delta.
\end{align}
In the $E_{11}$ integral in the definition of $H_\eps(x,\zeta,\nu)$ a turning point where $d>0$  is excised via the upper half plane and one where $d<0$ via the lower half plane.

In part (3) of Theorem \ref{M} one must correctly redefine the function  $\theta_1$ that appears in the statement of Proposition \ref{44l}, and the functions $\bar\theta_1^\pm$, $\theta_1^\pm$, $\bar\theta_2^+$, $\theta_2^+$, and constant  $\alpha(\eps,\zeta,\nu)$ that appear in the statement of Proposition \ref{88} and Corollary \ref{explicit}.

The function $\theta_1$ is defined for $x_{2*}<x<x_{1*}$ before analytic continuation around $x_{1*}$  by formulas \eqref{7a},\eqref{7b}, where $s$ is given by \eqref{a11z} on the real axis, and in the $k_1$ integration path the point $x_{2*}$ is excised via the upper half plane.

The functions $\theta_1^\pm$ are given  on  $x_{2*}<x<x_{1*}$ by the same function $\theta_1$  before analytic continuation around $x_{2*}$.
Their continuations and the corresponding exact solutions $\bar\theta_1^\pm$  are then constructed as in case I.  The functions $\theta_2^+$ and $\bar\theta_2^+$ are then defined as in case I.  The constant $\alpha(\eps,\zeta,\nu)$ is given by \eqref{4u}, where the contour starts at $0$ and encircles just $x_{2*}$.

\begin{prop}\label{M}
In case M let $\zeta\in \text{Class III}\setminus\{ ic_0\eta^{1/2}(0+),ic_0\eta^{1/2}(M), ic_0\eta^{1/2}(\infty)\}$.

i)Suppose there is only one turning point $x_*$ corresponding to $\zeta$ and that $\frac{d(c_0^2\eta)}{dx}(x_*)<0$.  Then the
asymptotic behavior of $\bar\theta_{1,G}$ on $[0,x_*-3\delta]$ is given by Proposition \ref{44l}.

ii.) Suppose there is only one turning point $x_*$ corresponding to $\zeta$ and that $\frac{d(c_0^2\eta)}{dx}(x_*)>0$.  Then the
asymptotic behavior of $\bar\theta_{1,G}$ on $[0,x_*-3\delta]$ is given by Proposition \ref{88} and Corollary \ref{explicit}.

iii.) Suppose there are two turning points $x_{2*}<x_{1*}$ with
\begin{align}\label{9bz}
\frac{d(c_0^2\eta)}{dx}(x_{1*})<0<\frac{d(c_0^2\eta)}{dx}(x_{2*}).
\end{align}
On $[x_{2*}+3\delta,x_{1*}-3\delta]$ the
asymptotic behavior of $\bar\theta_{1,G}$ is given by Proposition \ref{44l}, while on $[0,x_*-3\delta]$ the asymptotic behavior of  $\bar\theta_{1,G}$  is given by Proposition \ref{88} and Corollary \ref{explicit}.  In particular, $\bar\theta_{1,G}(0,\zeta,\nu,\eps)$ is given again by formula \eqref{4vr}, with $\alpha$ defined as explained above.

\end{prop}

\begin{proof}
\textbf{1. }Part (i) (resp., part (ii)) follows immediately from the fact that for $x<x_*$, $s^2=\zeta^2+c_0^2\eta>0$ (resp. $<0$) just as in case D (resp. case I).

\textbf{2. }In the proof of part (iii)  the functions $\theta_{1,M}$ and $\bar\theta_{1,M}$ are  defined near $x_{1*}$ as in case D, using the redefinition of $\theta_1$ described above. The analytic extension of $\bar\theta_{1,M}$ to
$[x_{2*}+3\delta,x_{1*}-3\delta]$ satisfies
\begin{align}\label{9bw}
|\bar\theta_{1,M}-\theta_1|\leq \frac{C}{\eps}|\theta_1|\text{ on }  [x_{2*}+3\delta,x_{1*}-3\delta].
\end{align}
by the argument using Theorem \ref{mpp} that gave \eqref{8b}.  The redefinition of the other exact solutions $\bar\theta_{21}$, $\bar\theta_{22}$, $\dots$,$\bar\theta_5$ needed for the matching argument near $x_{1*}$ is similar to the redefinition of  $\bar\theta_{1,M}$.
     An estimate like \eqref{8oz} is now used to show that
\begin{align}\label{9bq}
|\bar\theta_{1,G}-\bar\theta_{1,M}|\leq \frac{C}{\eps}|\bar\theta_{1,M}|\text{ on }[x_{2*}+3\delta,x_{1*}-3\delta].
\end{align}

\textbf{3. }The redefinition of the other exact solutions $\bar\theta_{1,M},\bar\theta_2,\dots,\bar\theta_5$ needed for the matching argument near $x_{2*}$ is similar to the redefinition of $\bar\theta_1^\pm$. In place of \eqref{4h} we can set
$x_2=x_{2*}+3\delta$ and use \eqref{9bw}, \eqref{9bq} to conclude
\begin{align}\label{pbp}
|\bar\theta_{1,G}-\theta_1|(x_2)\leq \frac{C}{\eps}|\theta_1|(x_2).
\end{align}
The rest of the proof follows as in case I.

\end{proof}

\begin{rem}\label{beh}
\textup{The results of this section show that in case D (resp. case I) the exact decaying solution of \eqref{a7} decays (resp. oscillates) to the left of the turning point, and oscillates  (resp. decays) to the right of the turning point.  In case M we have oscillation to the left of $x_{2*}$, followed  by decay, followed by oscillation to the right of $x_{1*}$.  In cases D and M, when $\Re\nu>0$ the oscillation on the unbounded interval occurs with an amplitude that decays exponentially to zero.}
\end{rem}

\section{The Instability Theorem}\label{instab}
\emph{\quad}In this section we present, with some simplifications and extensions, the main stability and instability results of \cite{Er3}.  In particular, we give the argument of \cite{Er3} for locating unstable zeros of
\begin{align}\label{n1}
V(\tau,\eps)=\tau b_1(\tau,\eps)+i\eps b_2(\tau,\eps)-\theta(0,\tau,\eps)\cdot (\tau h_t+ i\eps h_y),
\end{align}
where $\theta$ is the exact decaying solution of \eqref{a7} on $[0,\infty)$, and the other quantities appearing in \eqref{n1} are defined in section \ref{stabilityfn}.  With $\tau=\zeta\eps+\nu$ we rewrite this as
\begin{align}\label{n2}
\begin{split}
&V(\tau,\eps)=\eps L(\eps,\zeta,\nu),\\
&L(\eps,\zeta,\nu)=\zeta b_1+ib_2-\theta(0,\zeta,\nu,\eps)\cdot(\zeta h_t+ih_y)+\eps^{-1}\nu[b_1-\theta(0,\zeta,\nu,\eps)\cdot  h_t].
\end{split}
\end{align}
For $\Re\zeta\geq 0$ and all three types of profiles, the integrals $b_1$, $b_2$ are shown in \cite{Er3}, p.57-64 to approach zero as $\eps\to \infty$.  This part of the argument, which is complicated since the integrands involve $\theta$,  can be eliminated  if one uses instead the following simpler form of the stability function derived in \cite{CJLW}, Definition 4.7, and shown there to equal $V$:
\begin{align}\label{n3}
V(\tau,\eps)=\theta(0,\zeta,\nu,\eps)\cdot w(0+)-\theta(0,\zeta,\nu,\eps)\cdot(\tau h_t+i\eps h_y).
\end{align}
Using \eqref{n3}, one finds by inspection
\begin{align}\label{n4}
|L(\eps,\zeta,\nu)-L_{a}(\eps,\zeta,\nu)|\leq C_K/\eps \text{ as }\eps\to\infty,\text{ where }L_{a}:=-\theta(0,\zeta,\nu,\eps)\cdot[\zeta h_t+ih_y],
\end{align}
 and the constant $C_K$ is uniform for $\zeta$ in any compact subset $K$.\footnote{The rate of convergence to $0$ for $L-L_{a,}$ is computed in \cite{Er3} for profiles I, D, or M to be $O(\eps^{-1/2})$ with no discussion of uniformity with respect to $\zeta$.   By \eqref{n3} the rate $C_K/\eps$ holds, even for profiles not of type I, D, or M.}

The next proposition, which is the main step in the proof of the theorem that follows, provides explicit formulas for the approximate stability function $L_{a}$:
\begin{prop}\label{n4a}
Let $t_1(\zeta)$ and $t_2(\zeta)$ denote the first two columns of the matrix $T(0,\zeta)$ \eqref{a12}.
Under Assumptions \ref{thermo}, \ref{profile}, and \ref{rate} we have:

1)For $\zeta\notin\text{Class III}$
\begin{align}\label{n4b}
|L_a(\eps,\zeta,\nu)-L_1(\zeta)|\leq C_K/\eps, \text{ where } L_1(\zeta):=-t_1(\zeta)\cdot (\zeta h_t+i h_y),
\end{align}
 and the jump terms $h_t$ and $h_y$ are given in section \ref{stabilityfn}.  The constant $C_K$ is uniform for $\zeta$ in any compact subset $K\subset (\text{Class III})^c$.  This result holds even for profiles that are not of type $I$, $D$, or $M$.

2) For  profiles of type D and $\zeta\in\text{Class III}_+\setminus\{ ic_0\eta^{1/2}(0+),ic_0\eta^{1/2}(\infty)\}$, we have \begin{align}\label{n4bz}
|L_a(\eps,\zeta,\nu)-L_1(\zeta)|\leq C_\zeta/\eps.
\end{align}

3)For profiles of type I and $\zeta\in\text{Class III}_+\setminus \{ ic_0\eta^{1/2}(0+), ic_0\eta^{1/2}(\infty)\}$, we have
\begin{align}\label{n4c}
\left|L_a(\eps,\zeta,\nu)-[L_1(\zeta)+\alpha(\eps,\zeta,\nu)L_2(\zeta)]\right|\leq C_\zeta/\eps, \text{ where }L_2(\zeta):=-t_2(\zeta)\cdot (\zeta h_t+i h_y),
\end{align}
and $\alpha$ is given in \eqref{4u}.

4)For profiles of type M and $\zeta\in\text{Class III}_+\setminus \{ ic_0\eta^{1/2}(x_M),ic_0\eta^{1/2}(0+), ic_0\eta^{1/2}(\infty)\}$,   in the cases described by parts (i), (ii), and (iii) of Proposition \ref{M}, $L_a$ satisfies, respectively, \eqref{n4bz}, \eqref{n4c}, \eqref{n4c}.

In parts (2) and (3) and in the three subcases of part (4), the constant $C_\zeta$ can be chosen uniformly for $\zeta$ in a compact subset of the specified set of $\text{Class III}_+$ values.

\end{prop}

\begin{proof}
The result follows immediately from \eqref{n4} and the formulas for $\theta(0,\zeta,\nu,\eps)$ given  in Corollary \ref{notIII}, Corollary \ref{8qz}, Corollary \ref{explicit}, and Proposition \ref{M}, part (iii).
\end{proof}

The idea now is to study $V$ for large $\eps$ by studying the functions $L_1(\zeta)$ and $L_1(\zeta)+\alpha(\eps,\zeta,\nu)L_2(\zeta)$, which are given explicitly in \cite{Er3}.  The functions $L_1$ \eqref{n5} and $L_2$ \eqref{n6} are computed from the expressions for $T(0,\zeta)$, $h_t$, and $h_y$ already given, and $\alpha$ is readily computed using the expression for $E_{11}$ \eqref{E11} by letting the radius of the circular part of the contour $C$ in \eqref{4u} shrink to zero.  Writing $\zeta=i\zeta_i$, one obtains

\begin{align}\label{n7}
\begin{split}
&\alpha(\eps,\zeta,\nu)=e^{\beta(\eps,\zeta,\nu)}, \text{ where }\\
&\beta(\eps,\zeta,\nu)=\frac{\pi i}{2}-i\eps\beta_1(\zeta_i)+\beta_2(\zeta_i)-\nu\beta_3(\zeta_i),
\end{split}
\end{align}
and
\begin{align}\label{n8}
\begin{split}
&\beta_1=\int^{x_*(\zeta_i)}_0 \frac{2\kappa |s|}{\eta u} dx\\
&\beta_2=\zeta_i\int^{x_*(\zeta_i)}_0 \frac{\kappa}{\eta |s|}\left(\frac{1}{1-\eta}\frac{d\eta}{dx}-\frac{v p_S\sigma r}{Tu}+\frac{2u\sigma r}{(1-\eta)(u^2+\zeta_i^2)}-\frac{v\sigma r_v}{u}\right)dx\\
&\beta_3=\zeta_i\int^{x_*(\zeta_i)}_0\frac{2\kappa}{\eta|s| u}dx, \;\zeta=i\zeta_i.
\end{split}
\end{align}
Observe that $\alpha$ is periodic in $\eps$ and that the $\beta_i$ are all real, with $\beta_1$ and $\beta_3$ positive while the sign of $\beta_2$ depends on the profile $w$.

\begin{rem}\label{L1}
\textup{By Assumption \ref{vN} the stability function for the von Neumann step-shock, $L_{vN}(\zeta)$, is nonvanishing in $\Re\zeta\geq 0$.  Stability functions for step-shocks were  defined and analyzed in \cite{Er4}.  Remarkably, it turns out that
\begin{align}\label{n9}
L_{vN}(\zeta)=L_1(\zeta),
\end{align}
and the analysis of \cite{Er4} shows that since $L_1$ is nonvanishing in $\Re\zeta\geq 0$, in fact
\begin{align}\label{n9z}
L_1(\zeta)>0\text{ for }\zeta_i> c_0\eta^{1/2}(0+)
\end{align}
and increases monotonically with $\zeta_i$.  The function in \eqref{n9} is a nonvanishing multiple of the stability determinant for step-shocks defined in \cite{M}.}

\end{rem}

\begin{theo}\label{instability}

Let $\tau=\zeta\eps+\nu$, where $|\nu|\leq R$.
Under Assumptions \ref{thermo}, \ref{profile}, and \ref{rate} we have:

a)For $\zeta\in \{\Re\zeta\geq 0\}\setminus\text{Class III}$  there exists a positive constant $\eps(\zeta,R)$ such that
  $V(\tau,\eps)\neq 0$ for $\eps\geq \eps(\zeta,R)$.
This result holds even for profiles that are not of type $I$, $D$, or $M$.

b) For a  profile of type D and $\zeta\in\text{Class III}_+\setminus\{ ic_0\eta^{1/2}(0+),ic_0\eta^{1/2}(\infty)\}$, there exists a positive constant $\eps(\zeta,R)$ such that  $V(\tau,\eps)\neq 0$ for $\eps\geq \eps(\zeta,R)$.

c)For a profile of type I, suppose that for $i\zeta_i\in\text{Class III}_+\setminus \{ ic_0\eta^{1/2}(0+), ic_0\eta^{1/2}(\infty)\}$ one has the inequality
\begin{align}\label{n10}
|L_2(i\zeta_i)|\exp[\beta_2(\zeta_i)]<L_1(i\zeta_i).
\end{align}
Then there exists a positive constant $\eps(\zeta_i,R)$ such that  $V(i\zeta_i\eps+\nu,\eps)\neq 0$ for $\eps\geq \eps(\zeta_i,R)$.

d)For a profile of type I, suppose that  for some $i\zeta_i\in\text{Class III}_+\setminus \{ ic_0\eta^{1/2}(0+), ic_0\eta^{1/2}(\infty)\}$ one has
\begin{align}\label{n10z}
|L_2(i\zeta_i)|\exp[\beta_2(\zeta_i)]>L_1(i\zeta_i).
\end{align}
Let $\nu$ satisfying $|\nu|\leq R$ be an arbitrary point on the vertical line defined by
\begin{align}\label{n12z}
L_1=|L_2|\exp[\beta_2-\Re(\nu)\beta_3],
\end{align}
and suppose now that $R\geq \Re\nu$ for $\Re \nu$ as in \eqref{n12z}.  For any $\delta>0$  there exists a positive constant $\eps(\zeta_i,R,\delta)$ such that for periodically distributed $\eps\geq \eps(\zeta_i,R,\delta)$ satisfying
\begin{align}\label{n12}
\begin{split}
&\eps=[(2n-\frac{1}{2})\pi-\Im(\nu)\beta_3]/\beta_1, \;L_2>0\\
&\eps=[(2n+\frac{1}{2})\pi-\Im(\nu)\beta_3]/\beta_1, \;L_2<0,
\end{split}
\end{align}
$V(i\zeta_i\eps+\nu,\eps)$ has a zero within a distance $\delta$ of $\nu$.

(e) For a profile of type M, suppose that $i\zeta_i\in\text{Class III}_+\setminus \{ ic_0\eta^{1/2}(0+),ic_0\eta^{1/2}(x_M), ic_0\eta^{1/2}(\infty)\}$. If there is a single turning point $x_*$ and $d=\frac{d(c_0^2\eta)}{dx}(x_*)<0$, the conclusion of part (b) holds for $\zeta=i\zeta_i$.

Suppose there is a single turning point $x_*$ and $d>0$. If inequality \eqref{n10} holds, then the conclusion of part (c) holds, while  if inequality \eqref{n10z} holds, then the conclusion of part (d) holds.

 Suppose there are two turning points $x_{2_*}<x_{1*}$.  If inequality \eqref{n10} holds, then the conclusion of part (c) holds, while  if inequality \eqref{n10z} holds, then the conclusion of part (d) holds. Recall that when there are two turning points, the contour in the integral defining $\alpha$ encloses only $x_{2*}$.\\

The constants $\eps(\zeta,R,\delta)$ (resp. $\eps(\zeta_i,R,\delta)$) above can be chosen uniformly for $\zeta$ (resp. $\zeta_i$) in compact subsets of the sets of values described in the respective cases (or subcases) listed above.

%e) Statements (c) and (d) hold for profiles of type M, except that $\zeta_i=c_0\eta^{1/2}(x_M)$ needs to be excluded as well.

%\footnote{In \cite{Er3} the value $\zeta_i=c_0\eta^{1/2}(x_M)$ is \emph{not} excluded, as a consequence of the turning point %analysis done there near $x_M$.}

\end{theo}

\begin{proof}
\textbf{1. }First observe that as a consequence of \eqref{n4}, the estimates \eqref{n4b}, \eqref{n4bz} hold with $L_a$ replaced by $L$.  Parts (a) and (b) then follow immediately from the nonvanishing of $L_1(\zeta)$ and parts (1) and (2) of Proposition \ref{n4a}.

\textbf{2. }From \eqref{n7} we have
\begin{align}\label{n9y}
|\alpha|=\exp(\beta_2-\Re(\nu)\beta_3),\;\;\beta_3>0.
\end{align}
The inequality \eqref{n10} thus implies that $L_1(i\zeta_i)+\alpha(\eps,i\zeta_i,\nu) L_2(i\zeta_i)$ does not vanish for $\Re\nu\geq 0$.
Part (c) now follows from \eqref{n4} and part (3) of Proposition \ref{n4a}.

\textbf{3. }When the inequality \eqref{n10z} holds, \eqref{n7} implies that the analytic function of $\nu$ given by
\begin{align}\label{n13}
V_a(\nu;\zeta_i,\eps):=L_1(i\zeta_i)+\alpha(\eps,i\zeta_i,\nu)L_2(i\zeta_i)
\end{align}
has zeros on the vertical line \eqref{n12z} at the $\eps$ values given by \eqref{n12}. The nonvanishing of $\beta_3$ and $L_2(i\zeta_i)$ implies these zeros are simple.  Observe also that
\begin{align}\label{n14}
|\partial_\nu V_a(\nu;\zeta_i,\eps)|=e^{\beta_2-\Re(\nu)\beta_3}\beta_3|L_2|\text{ is independent of }\eps.
\end{align}
Fix an arbitrarily small $\delta>0$ and let $C_\delta$ be a circular contour centered at $\nu$ of radius $\delta$.  For $\eps$ large enough satisfying \eqref{n12},  part (3) of Proposition \ref{n4a} and \eqref{n14} imply
\begin{align}\label{n15}
|V(i\zeta_i\eps+\nu,\eps)-V_a(\nu;\zeta_i,\eps)|<\frac{C_{\zeta_i}}{\eps}<|V_a(\nu;\zeta_i,\eps)|\text{ on }C_\delta.
\end{align}
Part (d) now follows from Rouch\'e's Theorem.  The subcases of part (e) follow in the same way from part (4) of Proposition \ref{n4a}.

\textbf{4. }The statements about uniform choices of $\eps$ are direct consequences of the corresponding statements in Proposition \ref{n4a}.

\end{proof}

\noindent\textbf{Applications. }

1) We refer to \cite{Er3} for a detailed discussion of the consequences of Theorem \ref{instability}.  For one reaction $A\to B$ detonations with the Arrhenius rate law \eqref{rate2}, Erpenbeck first identifies ranges of various physical parameters for which profiles of type I, D, or M do occur.  Then he uses Theorem \ref{instability} to identify ranges of the parameters for which unstable zeros of $V$ are actually present in cases I and D.
Many of these results are also reported Chapter 6 of the book \cite{FD}.

2) Starting with  an open interval of $\zeta_i$ values satisfying \eqref{n10z}, one obtains from \eqref{n12}, say when $\Im(\nu)=0$,  corresponding $\eps$ intervals of unstable wavenumbers.  For $n$ sufficiently large these intervals overlap and one finds that $\emph{all}$ wavenumbers above a certain cutoff (depending on the interval of $\zeta$ values) are unstable.

%In order conclude from Theorem \ref

3) Most of the conclusions of \cite{Er3} for type M profiles do not rely on his analysis of the turning point problem for $\zeta_{i,M}:=c_0\eta^{1/2}(x_M)$, where $x_M$ is the location of the maximum. In particular, by applying part (e) of Theorem \ref{instability} to values of $\zeta_i$ near $\zeta_{i,M}$, but not equal to it, Erpenbeck derives the algebraic condition
\begin{align}\label{n16}
K(\zeta_{i,M}):=\left(\frac{1}{1-\eta}\frac{d\eta}{dx}-\frac{v p_S\sigma r}{Tu}+\frac{2u\sigma r}{(1-\eta)(u^2+\zeta_{i,M}^2)}-\frac{v\sigma r_v}{u}\right)(x_M)>0
\end{align}
as a sufficient condition for instability for profiles of type M.
The inequality \eqref{n16} implies the existence of unstable zeros of $V(i\zeta_i\eps+\nu,\eps)$ for $\zeta_i$ near $\zeta_{i,M}$. A separate  criterion is derived in \cite{Er3}, p.96, for $V$ to have unstable zeros exactly at the special value $\zeta_{i,M}$.\footnote{For $\zeta_{i,M}$ a formula like \eqref{4vr} is derived in \cite{Er3}, but with a completely different $\alpha$.}This condition cannot be derived from Theorem \ref{instability}.

\appendix
\section{Appendix}\label{app}

%\subsection{The tracking lemma}\label{s:track}
%
%Consider an approximately block-diagonal system
%\begin{equation}
%W'= \bp M_1 & 0 \\ 0 & M_2 \ep(x,p) W + \delta(x,p) \Theta(x,p) W,
%\label{blockdiag}
%\end{equation}
%where $\Theta$ is a uniformly bounded matrix, $\delta(x)$ scalar,
%and $p$ a vector of parameters,
%satisfying a pointwise spectral gap condition
%\begin{equation} \label{gap}
%\min \sigma(\Re M_1^p)- \max \sigma(\Re M_2^p)
%\ge \eta(x) >0
%\, \text{\rm for all } x.
%\end{equation}
%(Here as usual $\Re N:= \frac{1}{2}(N+N^*)$ denotes the
%``real'', or symmetric part of $N$.)
%
%\begin{lemma}[\cite{MaZ3,PZ,Z1}] \label{reduction}
%Consider a system \eqref{blockdiag} under the gap assumption
%\eqref{gap}, with $\Theta^p$ uniformly bounded and
%$\eta\in L^1_{\rm loc}$.
%If $\sup (\delta/\eta)(x)$ is sufficiently small,
%then there exists a unique linear
%transformation $\Phi(x,p)$,
%possessing the same regularity with respect to $p$
%as do coefficients $M_j$ and $\delta\Theta$
%(as functions into $L^\infty(x)$),
%for which the graph $\{(Z_1, \Phi Z_1)\}$ is invariant under
%%the flow of
%\eqref{blockdiag}, and
%%satisfies
%\ba\label{ptwise}
%|\Phi^p(x)| \le C
%\int_{-\infty}^{x} e^{\int_y^x -\eta(z)dz} \delta(y)dy
%\le \sup_{(-\infty,x]} (\delta/\eta).
%\ea
%\end{lemma}
%
%\begin{proof}
%See the proof of Lemma A.4 together with Remark A.6 in \cite{Z1}.
%\end{proof}
%

\subsection{A variable-coefficient gap lemma}\label{s:vargap}

\emph{\quad} In this section we state and prove Lemma \ref{vargaplem}, which is our main tool for determining the asymptotic behavior as $\eps=\frac{1}{h}\to\infty$ of the decaying solution $\theta(x,\tau,\eps)$ of \eqref{a7} on $[a,\infty)$.
This lemma was introduced in \cite{Z2} and used there for the study of stability of ZND detonations with respect to high frequency one-dimensional perturbations.

Consider a first-order  system
\be\label{newsys}
V'=A^p(x;h)V:=M^p(x;h)V +\Theta^p(x;h)V\text{ on } x\ge 0,
\ee
with $V(x)\in \CC^N$,  depending on a parameter $p\in P\subset  \RR^m$ and a distinguished  small parameter $h>0$.

\bl[\cite{Z2}] \label{vargaplem}
Suppose there exist positive constants $\beta$, $C$ such that for all $x\geq 0$ and $p\in P$:
\be\label{udecay2}
|\Theta^p(x,h)|\le C h^2 e^{-\beta h x}\text{ and }
\ee
\be\label{neutral2}
\Re M^p(x,h)\ge -(h \delta^p(h) + C h e^{-\beta h x}),\text{ where }0\leq\delta^p(h)\le \delta_*<\beta
\ee
for all $h$.\footnote{Here $\Re M:=\frac{1}{2}(M+M^*)$.} Assume further that there exists a nonzero constant vector $V_*^p$ such that
\begin{align}\label{m6}
M^p(x,h)V^p_* = 0 \text{ for all }x\geq 0.
\end{align}
Then there exists an $h_0$ independent of $p\in P$ such that for $0\leq h<h_0$,
 there exists a solution
$V^p(x,h)$ of \eqref{newsys} defined on $x\ge 0$ satisfying
\begin{equation}
\label{Pdecay2new}
|( V^p(x)-V^p_*)|  \le C_1 h e^{- \delta_*h x}|V_*^p|
\quad
\text{\rm for } x\ge 0.
\end{equation}
\el

\begin{proof}
%Suppressing the superscript $p $,
\textbf{1. }We seek, equivalently, a
solution $V^p$ of the integral fixed-point equation
\begin{equation}
\begin{aligned}
\label{fixmap}
\CalT V(x)
&= V^p_*+  \int^x_{+\infty} \mathcal{F}^{y\to x} \Theta^p (y,h)V(y) dy ,
\end{aligned}
\end{equation}
where $\mathcal{F}^{y\to x}$ is the solution operator of
$V'=M^p V$ from $y$ to $x$; that is, $W(x):=\mathcal{F}^{y\to x}w(y)$ satisfies
\begin{align}\label{m1}
W_x=M^pW,\;\;W|_{x=y}=w(y).
\end{align}

\textbf{2. }We claim
\be\label{growth2}
\|\mathcal{F}^{y\to x}\|\le Ce^{  h \delta^p(h) (y-x)}
\le Ce^{  h \delta_* (y-x)}
\; \hbox{ \rm for } \; y>x.
\ee
Indeed, for $W(x)$ as in \eqref{m1} we have using \eqref{neutral2}
\begin{align}\label{m2}
\begin{split}
&d_x(|W|^2)=2\Re(W,W_x)=2(W,\Re M^pW)\geq -2(h \delta^p(h) + C h e^{-\beta h x})|W|^2\Rightarrow\\
&\qquad\qquad|W|_x/|W|\geq -(h \delta^p(h) + C h e^{-\beta h x}),
\end{split}
\end{align}
and integrating this inequality from $x$ to $y$  gives
\begin{align}\label{m3}
|W(x)|/|w(y)|\leq C_1 e^{h\delta^p(h)(y-x)}, \text{ where }C_1=\int^\infty_0 Che^{-\beta h x}dx.
\end{align}

\textbf{3. }For $h>0$ sufficiently small, this implies that
$\mathcal{T}$ is a contraction on $L^\infty[0,\infty)$.
For, applying \eqref{udecay2} and \eqref{growth2} we have (with $|V|_\infty=|V|_{L^\infty[0,\infty)}$)
\begin{equation}\label{con2}
\begin{aligned}
\left|\CalT V_1 - \CalT V_2 \right|(x)
&\le Ch^2 |V_1 - V_2|_\infty
\int^\infty_x e^{\delta_* h (y-x)} e^{-\beta h y} dy
\le  C' h |V_1 - V_2|_\infty e^{-\delta_* h x},
\end{aligned}
\end{equation}
which for $h$ sufficiently small is less than $\frac{1}{2} |V_1 - V_2|_\infty$.
By iteration, we thus obtain a solution
$V^p \in L^\infty [0,\infty)$ of $V = \CalT V$.
Further, taking $V_1=V^p$, $V_2=0$ in \eqref{con2} we obtain both
\begin{align}\label{m4}
\begin{split}
&|V^p-V^p_*|_\infty=|\cT V^p-\cT0|_\infty\leq \frac{1}{2}|V^p|_\infty\Rightarrow |V^p|_\infty\leq 2|V^p_*|\quad\text{ and }\\
&|V^p-V^p_*|(x)\leq C' h |V^p|_\infty e^{-\delta_* h x},
\end{split}
\end{align}
which together imply \eqref{Pdecay2new}.
\end{proof}

%CHANGED: added this rmk for completeness -KZ
\begin{rem}\label{fullgap}
\textup{
Strictly speaking, Lemma \ref{vargaplem} is a ``weak'' gap lemma
realizing only the fastest decaying mode of \eqref{newsys},
whereas the full gap lemma construction of \cite{GZ} (see, for example,
Cor. 2.4, \cite{GZ}, or Prop. 3.1 of \cite{ZH}) yields
intermediate modes as well.
However, a straightforward modification of \eqref{fixmap} following
that of Prop. 3.1, \cite{ZH} yields intermediate modes as well,
provided that $M^p$ splits into diagonal blocks $M^p_\pm$
such that
$\Re M^p_\pm(x,h)\gtrless  -(h \delta^p(h) + C h e^{-\beta h x})$.
 %where $0\leq\delta^p(h)\le \delta_*<\beta$.
}
\end{rem}

\begin{rem}\label{finitegap}
\textup{
%It is worth noting that, on an interval
On an interval $[a/h,b/h]$, corresponding in coordinates $\tilde x=hx$
to a finite interval $[a,b]$, essentially the same argument
as in the proof of Lemma \ref{vargaplem} applies with modified
%fixed-point
mapping
\begin{equation}
\begin{aligned}
\label{finitefixmap}
\CalT V(x)
&= V^p_*+  \int^x_{b/h} \mathcal{F}^{y\to x} \Theta^p (y,h)V(y) dy
\end{aligned}
\end{equation}
to yield an analogous result
$|( V^p(x)-V^p_*)|  \le C_1 h |V_*^p|$
for $ x\in [a/h,b/h]$,
assuming only uniform boundedness of the coefficient matrix $A^p$,
%CHANGED AGAIN: added more to make correct... -KZ
$|\Theta^p(x,h)|\le C h^2 $, and
$\Re M^p(x,h)\ge -Ch$,
or, for intermediate modes,
$\Re M^p_\pm(x,h)\gtrless  - C h$,
similarly as discussed in Remark \ref{fullgap}.
%CHANGED:slight changes for added clarity-KZ
%Comparing with Theorem \ref{mpp}(b), and
%rescaling $x\to hx$, we see that,
Rescaling $x\to hx$, and comparing with Theorem \ref{mpp}(b), we see that,
%ENDCHANGED
restricted to a finite closed real interval, the
Method of Parameters and the variable-coefficient
gap lemma constructions thus give essentially the same results
(here restricted to level $m=0$).
On an infinite interval, on the other hand,
it is readily checked that the
truncation error $|\Phi_m-\Phi|$ resulting from
the Method of Parameters construction described in \eqref{a13}
generically does not decay in $x$,
%CHANGED: added this..
even for $\eps \Phi(\eps, z)=
\eps \Phi_0 + \Phi_1$ with $\Phi_j$ constant (rather than
just asymptotically constant),
%ENDCHANGED
and so the method of parameters in general fails.\footnote{
Specifically, the righthand side of \eqref{5g} blows up as $z\to \infty$.}
}
\end{rem}
%ENDCHANGED

\subsection{ZND equations}
\emph{\quad} The ZND equations for the unknowns $(v,\mathbf{u},S,\lambda)$ (specific volume, particle velocity $\mathbf{u}=(u_x,u_y,u_z)$, entropy, and mass fraction of reactant) are

\begin{align}\label{ZND}
\begin{split}
&\partial_t v+\mathbf{u}\cdot \nabla v-v\nabla \cdot \mathbf{u}=0\\
&\partial_t \mathbf{u}+\mathbf{u}\cdot\nabla \mathbf{u}+v\nabla p =0\\
&\partial_t S+\mathbf{u}\cdot\nabla S=-r \Delta F /T:=\Phi\\
&\partial_t\lambda+\mathbf{u}\cdot\nabla \lambda =r,
\end{split}
\end{align}
where $p=p(v,S,\lambda)$ is pressure, $T$ is temperature, $\Delta F$ is the free energy increment, and $r(v,S,\lambda)$ is the reaction rate function.

\subsection{The stability function $V(\tau,\eps)$.}\label{stabilityfn}
\emph{\quad}The stability function defined in \cite{Er1} and used in $\cite{Er3}$ is given by
\begin{align}\label{V}
V(\tau,\eps)=\tau b_1(\tau,\eps)+i\eps b_2(\tau,\eps)-\theta(0,\tau,\eps)\cdot (\tau h_t+ i\eps h_y),
\end{align}
where (with $v'=\frac{dv}{dx}$ and $m=u/v$, a constant independent of $x$)
\begin{align}\label{V1}
g_t=-(v',u',0,S',\lambda')^t,\;\;g_y=(0,0,-vp',0,0)^t,\;\;b_j=-\int^\infty_0\theta(x,\tau,\eps)\cdot A_x^{-1}(x) g_j(x)dx \text{ for }j=1,2,
\end{align}
\begin{align}\label{V2}
h_t=\frac{v_--v_+}{v_-T_+\eta_+}\begin{pmatrix}2(1-\eta_+)g_+/m\\T_+\eta_++2(1-\eta_+)g_+\\0\\-m(v_--v_+)\eta_+\\0\end{pmatrix},
\end{align}
and $h_y$ the single has nonzero component $(h_y)_3=m(v_--v_+)$.   Here $v_\pm$, for example, are components of the states $w_\pm:=w(0_\pm)$ just to the right and left of the von Neumann shock, and
\begin{align}\label{V3}
g_+=T_+-\frac{1}{2}(v_--v_+)p_{S+}.
\end{align}
In section \ref{instability} we work with the simpler form of $V$ \eqref{n3} in which the integals $b_j$ do not appear.

The functions $L_1(\zeta)$ and $L_2(\zeta)$, appearing in Proposition \ref{n4a} and used to compute zeros of $V$, are given explicitly in \cite{Er3} as:

\begin{align}\label{n5}
\begin{split}
&L_1(\zeta)=-\frac{u_-(1-\chi_v)}{\eta_+}\left[\frac{\ell_+\zeta(\zeta+\kappa_+s_+)}{u_+u_-}+\eta_+\left(1-\frac{\zeta^2}{u_+u_-}\right)\right]\\
&\ell=2-(1-\eta)(1-\chi_v)v_-p_S/T,\;\chi_v=v_+/v_-,
\end{split}
\end{align}
where $\pm$ denotes evaluation at $0\pm$.

\begin{align}\label{n6}
L_2(\zeta)=-\frac{u_-(1-\chi_v)}{\eta_+}\left[\frac{\ell_+\zeta(\zeta-\kappa_+s_+)}{u_+u_-}+\eta_+\left(1-\frac{\zeta^2}{u_+u_-}\right)\right].
%&\ell=2-(1-\eta)(1-\chi_v)v_-p_S/T,\;\chi_v=v_+/v_-,
\end{align}

\subsection{Coefficients appearing in the linearized systems}\label{coefficients}

\emph{\quad}The matrix coefficients appearing in the reduced system \eqref{a5} are
\begin{align}\label{A1}
\begin{split}
&A_x=\begin{pmatrix}u&-v&0&0&0\\vp_v&u&0&vp_S&vp_\lambda\\0&0&u&0&0\\0&0&0&u&0\\0&0&0&0&u\end{pmatrix},\;A_y=\begin{pmatrix}0&0&-v&0&0\\0&0&0&0&0\\vp_v&0&0&vp_S&vp_\lambda\\0&0&0&0&0\\0&0&0&0&0\end{pmatrix}\\
&B=\begin{pmatrix}-u'&v'&0&0&0\\p'-v(c_0^2/v^2)'&u'&0&vp_S'&vp_\lambda'\\0&0&0&0&0\\-\Phi_v&S'&0&-\Phi_S&-\Phi_\lambda\\-r_v&\lambda'&0&-r_S&-r_\lambda\end{pmatrix},
\end{split}
\end{align}
where $(')$ denotes differentiation with respect to $x$ and $c_0^2=-v^2p_v(v,S,\lambda)$.   These matrices are obtained from the corresponding matrices in the unreduced $6\times 6$ system by deleting the fourth row and fourth column of each of the latter matrices.  In each of the unreduced matrices $A_y$ (coefficient of $\alpha$) and $B$, the fourth row and fourth column consist only of zeros, while the fourth row and fourth column of $A_x$ have the fourth component $u$ and all other components $0$.  The matrix coefficient of $\beta$ in the unreduced system is
\begin{align}\label{A2}
A_z=\begin{pmatrix}0&0&0&-v&0&0\\0&0&0&0&0&0\\0&0&0&0&0&0\\vp_v&0&0&0&vp_S&vp_v\\0&0&0&0&0&0\\0&0&0&0&0&0\end{pmatrix}.
\end{align}

The matrix $\Phi_0(x,\zeta)$ in the transposed system \eqref{a9} is computed in \cite{Er3}, p.112 to be
\begin{align}\label{phi0}
\Phi_0(x,\zeta)=\begin{pmatrix}-\frac{(1-\eta)\zeta}{\eta u}&-\frac{m\zeta}{\eta u}&-\frac{im}{1-\eta}&0&0\\-\frac{(1-\eta)\zeta}{\eta mu}&-\frac{(1-\eta)\zeta}{\eta u}&0&0&0\\\frac{i(1-\eta)}{\eta m}&\frac{i}{\eta}&\frac{\zeta}{u}&0&0\\\frac{(1-\eta)p_S\zeta}{\eta m^2 u}&\frac{(1-\eta)p_S\zeta}{\eta m u}&\frac{ip_S}{m}&\frac{\zeta}{u}&0\\\frac{(1-\eta)p_\lambda\zeta}{\eta m^2 u}&\frac{(1-\eta)p_\lambda\zeta}{\eta m u}&\frac{ip_\lambda}{m}&0&\frac{\zeta}{u}\end{pmatrix}.
\end{align}

The matrix $E=T^{-1}\Phi_1 T-T^{-1}\frac{dT}{dx}$ \eqref{a12a} has $E_{11}$ component given by
\begin{align}\label{E11}
\begin{split}
&E_{11}(x,\zeta,\nu)=-\frac{(1-\eta)vp_S}{2T\eta u}\sigma r-\frac{1-\eta}{2\eta u}(2\nu+v\sigma r_v)+\frac{2-\eta}{4\eta(1-\eta)}\frac{d\eta}{dx}+\\
&\frac{\kappa \zeta}{2\eta s}\left(\frac{1}{1-\eta}\frac{d\eta}{dx}-\frac{vp_S}{Tu}\sigma r-\frac{1}{u}(2\nu+v\sigma r_v)\right)+\frac{\kappa\zeta+s}{\zeta+\kappa s}\frac{\zeta}{\eta u s}\sigma r-\frac{1}{2}\frac{d \ln s}{dx},
\end{split}
\end{align}
where $\sigma$ is defined (\cite{FD},p.95) by
\begin{align}\label{E12}
\sigma=v(\partial p/\partial\lambda)_{e,v}/c_0^2  \;\;\;\;(e\text{ is specific internal energy)}.
%v p_\lambda=c_0^2\left[\sigma+\beta_0 \frac{\Delta F}{vc_{p0}}\right].
\end{align}
%In \eqref{E12} $\beta_0$ is the expansion coefficient $\frac{\partial v}{\partial T}$ at fixed $(p,\lambda)$, and $c_{p0}$ is specific heat at %constant pressure.
The formula for $E_{22}$ is the same as \eqref{E11}, except that $s$ is replaced by $-s$.


\begin{thebibliography}{GMWZ2}

{\footnotesize

\bibitem[BZ1]{BZ1} Barker, B., and Zumbrun, K.,
\emph{A numerical investigation of stability
of ZND detonations for Majda's model,}
preprint (2010).

\bibitem[BZ2]{BZ2} Barker, B., and Zumbrun, K.,
{\it Numerical stability analysis of ZND detonations,}
in preparation.
%TODO: update.

\bibitem[CL]{CL} Coddington, E.A.,  and Levinson, N.,
{\it Theory of Ordinary Differential equations,}
McGraw--Hill Book Company, Inc., New York (1955).
%For directly applicat bale version of the theorm, see problemm 29, p. 104.

%\bibitem[Co]{Co} W. A. Coppel,
%{\it Stability and asymptotic behavior of differential equations},
%D.C. Heath and Co., Boston, MA (1965).

\bibitem[CJLW]{CJLW} Costanzino, N.,  Jenssen, H. K., Lyng,  G., and  Williams, M.,
{\it Existence and stability of curved multidimensional detonation fronts,}
Indiana Univ. Math. J.  56 (2007), no. 3, 1405--1461.

%\bibitem[CF]{CF} R. Courant and K.O. Friedrichs,
%{\it Supersonic flow and shock waves,}
%Springer--Verlag, New York (1976) xvi+464 pp.

\bibitem [Er1]{Er1}  Erpenbeck, J. J.,
{\it Stability of steady-state equilibrium detonations,}
Phys. Fluids 5 (1962),
604--614.

\bibitem [Er2]{Er2}  Erpenbeck, J. J.,
{\it Stability of idealized one-reaction detonations,}
Phys. Fluids 7 (1964). 684--696.

\bibitem [Er3]{Er3}  Erpenbeck, J. J.,
{\it Detonation stability for disturbances of small transverse wave length},
Phys. Fluids 9 (1966) 1293--1306;  \emph{Stability of detonations for disturbances of small transverse wavelength}, Los Alamos Preprint, LA-3306, (1965), 140 pages.

\bibitem [Er4]{Er4}  Erpenbeck, J. J.,
{\it Stability of step shocks.} Phys. Fluids 5 (1962) no. 10, 1181--1187.


\bibitem[FD]{FD}
Fickett, W. and Davis, W., \emph{Detonation: Theory and
Experiment}, Univ. California Press, Berkeley, 1979.

\bibitem[GZ]{GZ}
Gardner, R. A., and Zumbrun, K.,
{\it The gap lemma and geometric criteria for instability of viscous shock
profiles,} Comm. Pure Appl. Math. 51 (1998), no. 7, 797--855.

%\bibitem [Er5]{Er5} J. J. Erpenbeck,
%{\it Nonlinear theory of unstable
%one--dimensional detonations,}
%Phys. Fluids 10 (1967) No. 2, 274--289.

\bibitem[HuZ]{HuZ}  Humpherys, J. and  Zumbrun, K.,
{\it Numerical stability analysis of detonation waves in ZND},
preprint (2010).

%\bibitem[HuZ2]{HuZ2} J. Humpherys and K. Zumbrun,
%{\it An efficient shooting algorithm for Evans function calculations
%in large systems,} Phys. D  220  (2006),  no. 2, 116--126.

\bibitem[JLW]{JLW}  Jenssen, H.K.,  Lyng, G.,  and Williams, M.,
{\it Equivalence of low-frequency stability conditions for
  multidimensional detonations in three models of combustion,}
Indiana Univ. Math. J. 54 (2005) 1--64.

\bibitem[LS]{LS}
  Lee, H.I. and Stewart, D.S.,
 {\it Calculation of linear detonation instability: one-dimensional
    instability of plane detonation,}
  J. Fluid Mech., 216 (1990) 103--132.


\bibitem[M]{M}
Majda, A., \emph{The stability of multidimensional shock fronts},
Mem. Amer. Math. Soc. No. 275, AMS, Providence, 1983.

\bibitem[Sh]{Sh}
Short, M., \emph{Multidimensional linear stability of a detonation
wave at high activation energy}, SIAM J. Appl. Math., 57,1997,
307-326.

\bibitem[SS]{SS}
Short, M. and  Stewart, D.S., \emph{The multidimensional stability
of weak heat-release detonations}, J. Fluid Mech., 382, 1999,
109-135.

\bibitem[TZ]{TZ}
Texier, B.,  and Zumbrun, K.,
\emph{Transition to longitudinal instability of detonation waves is
generically associated with Hopf bifurcation to time-periodic
galloping solutions,} to appear, Comm. Math. Phys.

\bibitem[Z1]{Z1}
Zumbrun, K.,
\emph {Stability of detonation waves in the ZND limit,}
to appear, Arch. for Rat. Mech. Anal.

\bibitem[Z2]{Z2} Zumbrun, K.,
{\it High-frequency asymptotics and 1-D stability of ZND detonations
in the small-heat release and high-overdrive limits},
preprint (2010).

\bibitem[Z3]{Z3}
Zumbrun, K., \emph{Multidimensional stability of planar viscous
shock waves}, Advances in the theory of shock waves, 304-516.
Progress in Nonlinear PDE, 47, Birkh\"auser, Boston, 2001.

\bibitem[ZH]{ZH} Zumbrun, K., and Howard, H.,
{\it Pointwise semigroup methods for stability of viscous shock waves,}
Indiana University Journal Vol. 47 (1998), pp. 727-841.

% preprint (2009).
}

\end{thebibliography}
\end{document}